\documentclass[a4paper,11pt]{article}
\pdfoutput=1 
\usepackage{jheppub}
\usepackage[T1]{fontenc} 
\usepackage[all]{xy}
\usepackage{rotating}
\usepackage{float}
\usepackage{tikz}
\usetikzlibrary{calc,decorations.markings}

\let\a=\alpha \let\b=\beta   
    
    \let\p=\pi

\def\a{\alpha}
\def\b{\beta}

\def\CD{{\cal D}}

\def\CF{{\cal F}}

\def\CH{{\cal H}}
\def\CI{{\cal I}}

\def\CL{{\cal L}}
\def\CM{{\cal M}}
\def\CN{{\cal N}}
\def\CO{{\cal O}}

\def\CR{{\cal R}}
\def\CS{{\cal S}}
\def\CT{{\cal T}}

\def\CW{{\cal W}}

\def\CZ{{\cal Z}}

\def\beq#1\eeq{\begin{align}#1\end{align}}

\makeatletter
\newcommand*{\rom}[1]{\expandafter\romannumeral #1}
\makeatother

\title{\boldmath 3D  bulk   field theories 
for 2D  non-unitary $\CN=1$ supersymmetric  minimal models}

\abstract{We propose bulk 3D $\CN=4$ rank-0 superconformal field theories, which are related to 2D $\CN=1$ supersymmetric minimal models, $SM(2, \cdot)$ and $SM(3,\cdot)$, via recently discovered non-unitary bulk-boundary correspondence.  The correspondence relates a 3D $\CN=4$ rank-0 superconformal field theory to  2D chiral rational conformal field theories. A topologically twisted theory of the rank-0 SCFT supports the rational chiral algebra  at the boundary upon a proper choice of boundary condition. We test the proposal  by checking several non-trivial dictionaries of the  correspondence. }

\author{Seungjoo Baek} 
\author{and Dongmin Gang}

\affiliation{
Department of Physics and Astronomy $\&$ Center for Theoretical Physics,
\\
Seoul National University, 1 Gwanak-ro, Seoul 08826, Korea}

\emailAdd{seungjoobaek@snu.ac.kr}
\emailAdd{arima275@snu.ac.kr}

\begin{document} 
\maketitle
\flushbottom


\section{Introduction}
2D rational conformal field theories (2D RCFTs) and 3D topological field theories (3D TQFTs) are the most recurring themes in theoretical and mathematical physics. 
They describe universal behaviors of  critical phenomena in 2D statistical models and (2+1)D topological orders respectively. Mathematically, RCFTs have underlying rational chiral algebra, a.k.a rational vertex operator algebra (rational VOA). 
 Both RCFTs and TQFTs share a rigid mathematical structure called modular tensor category (MTC), which defines a framed topological invariant of 3-manifolds and  knots \cite{zbMATH04092352,moore1989classical,turaev1992modular}. Some  substructures of the MTC  are found to be ubiquitous in quantum field thories in various space-time dimensions and play a key role in recent developments of generalized symmetry, see $e.g.$ \cite{Gaiotto:2014kfa,Bhardwaj:2017xup,Chang:2018iay,Heidenreich:2021xpr,Koide:2021zxj,Choi:2021kmx,Kaidi:2021xfk}. Based on the common MTC structure, the two topics are beautifully connected via so-called bulk-boundary correspondence \cite{zbMATH04092352,moore1989lectures}. 
 
 There also exist non-unitary  RCFTs and TQFTs. Non-unitary RCFTs arise from 2D statistical models with imaginary parameters \cite{fisher1978yang,cardy1985conformal}. Non-unitary chiral algebra also appears as a BPS subsector of  supersymmetric quantum field theories (SQFTs) in higher  $(D>2)$ dimensions \cite{Beem:2013sza,Beem:2014kka,Feigin:2018bkf,Cheng:2018vpl,Costello:2018fnz,Costello:2020ndc,Cheng:2022rqr}. For most cases, however, the non-unitary chiral algebra from SQFTs are irrational \cite{Song:2017oew,Beem:2017ooy,Arakawa:2017fdq,Xie:2019vzr,Gukov:2020lqm,Creutzig:2021ext}. Recently, a physical realization of 3D non-unitary (semi-simple and finite) TQFTs is proposed through full topological twistings of an exotic class of 3D $\CN=4$  superconformal field theories (SCFTs) called rank-0 SCFTs \cite{Gang:2018huc,Gang:2021hrd}.  
\ Being rank-0 means that there are no Coulomb or Higgs branch operators in the theory and the property turned out to be crucial to support  rational chiral algebra at the boundary \cite{Costello:2018swh,Beem:2023dub,Ferrari:2023fez}.
 Using the physical realization of non-unitary TQFTs, the bulk-boundary correspondence has been extended to the non-unitary cases in the following way:
 \begin{align}
 \fbox{3D  rank-0 SCFT } \xrightarrow{  \rm top'l \; twistings  } \fbox{3D non-unitary TQFTs}  \xrightarrow{  \textrm{at boundary}} \fbox{2D chiral RCFTs}\label{non-unitary bulk-boundary correspondence}
 \end{align}
 Refer to \cite{Gang:2023rei,Gang:2022kpe,Gang:2023ggt,Ferrari:2023fez,Dedushenko:2018bpp,Dedushenko:2023cvd} for the recent developments along the direction.

Using the underlying rigid mathematical structures, there has been efforts to classifiy  RCFTs. Famously, unitary 2D CFTs with central charge $c<1$ are all classified and they form a series of RCFTs called Virasoro minimal model.  The Virasoro minimal model can be further generalized to include non-unitary ones and it will be denoted by  $M(p,q)$ with two coprime integers $p$ and $q$ ($2\leq p<q $).  They have the Virasoro algera  with certain rational values  of $c$  as the underlying chiral algebra.  The $M(p, q)$ is unitary if and only if $|p-q|=1$. For Lee-Yang series of the non-unitary minimal model, $M(2,q)$ with $q\in 2\mathbb{Z}_{\geq 2}+1$, the corresponding bulk rank-0 SCFTs were proposed recently \cite{Gang:2023rei}. There also exist supersymmetric version of minimal models, supersymmetric minimal model  $SM(p,p')$ labeled by two integers subjected to the conditions in \eqref{SM(p,p')}. The underlying rational chiral algebra is 2D $\CN=1$ super-Virasoro algebra with certain rational values of $c$. The $SM(p,p')$ is unitary if and only if $|p-p'|= 2$. In this paper, we propose the bulk 3D rank-0 SCFTs corresponding to the non-untary 2D supersymmeric minimal models with $p=2$ and $p=3$.

The rest of this paper is organized as follows. In section \ref{sec : main}, we first review some basic aspects of supersymmetric minimal models and non-unitary bulk-boundary correspondence. Basic dictionaries of the correspondence are summarized in Table \ref{Table : Dictionaries}. Then, we propose the bulk $\CN=4$ rank-0 SCFTs corresponding to supersymmetric minimal models $SM(p,p') $ with $p=2,3$. For $p=2$, we present 3 distinct UV gauge theory descriptions which are claimed to be IR equivalent modulo a decoupled invertible spin-TQFT $SO(1)_1$, i.e. Ising spin-TQFT. The theories have $\CN=4$ (resp. $\CN=5$) supersymmetry in the IR for  $p=3$ (resp. $p=2$). In appendix \ref{appendix}, we also present new UV description of bulk field theories for minimal models $M(p,q)$ with $p=2$, which are claimed to be IR equivalent to the theories in \cite{Gang:2023rei} modulo a decoupled invertible spin TQFT $SO(1)_1$, Ising spin-TQFT.  

\section{Bulk dual $\CN=4$ rank-0 SCFTs of $\CN=1$ minimal models} \label{sec : main}
\subsection{Supersymmetric minimal model $SM(p,p')$ and rank-0 SCFT $\CT_{(p,p')}$}
Here we review basic aspects of the minimal model $SM(p,p')$ and the non-unitary bulk-boundary correspondence. 

\paragraph{Supersymmetric minimal model $SM(p,p')$} The $\CN=1$ minimal model $SM(p,p')$ is labeled by two integers, $p$ and $p'$, satisfying
\begin{align}
2 \leq p<p', \;\; p'- p \in 2\mathbb{Z}\textrm{ and gcd}\left( p, \frac{(p'-p)}2\right) =1\;. \label{SM(p,p')}
\end{align}
Primaries $\CO_{(s,t)}$ are labeled by two integers, $1\leq s \leq p-1$ and $1\leq t \leq p'-1$, with an equivalence relation $\CO_{(s,t)} = \CO_{(p-s,p'-t)}$. $\CO_{(s,t)=(1,1)}$ is  the identity operator.   
 Central charge $c$ and conformal dimensions $h$ of the primaries are
\begin{align}
\begin{split}
&c = \frac{3}2 \left(1- \frac{2(p'-p)^2}{p p'} \right)\;,
\\
& h_{(s,t)} = \begin{cases} &\frac{(p' s- pt)^2 - (p-p')^2}{8 p p'}\;, \quad s-t \in 2\mathbb{Z} \;\; ( \textrm{NS sector})
\\
&\frac{(p' s- pt)^2 - (p-p')^2}{8 p p'} + \frac{1}{16}\;, \quad \quad s-t \in 2\mathbb{Z}+1\;\; ( \textrm{R sector})
\end{cases} \label{c and h}
\end{split}
\end{align}
\\
Conformal characters for NS sectors are ($s-t \in 2\mathbb{Z}$) 
\begin{equation}
\chi_{(s,t)}= q^{h_{(s,t)}- \frac{c}{24}}\frac{(-q^{\frac{1}{2}};q)_{\infty}}{(q;q)_{\infty}}\sum_{l\in\mathbb{Z}}\left(q^{\frac{l(lpp'+sp'-tp)}{2}}-q^{\frac{(lp+s)(lp'+t)}{2}}\right)\,.  \label{SM-characters}
\end{equation}
We use following $q$-Pochhammer symbols
\begin{align}
\begin{split}
&(x;q)_k := \prod_{n=0}^{k-1}(1-q^n x), \quad (x;q)_{\infty} :=\prod_{n=0}^{\infty} (1-q^n x),
\\
&(q)_k:= (q;q)_k =\prod_{n=1}^k (1-q^n), \quad (q)_\infty := (q;q)_{\infty} = \prod_{n=1}^{\infty} (1-q^n)\;.
\end{split}
\end{align}
Modular S-matrix for the ${\rm NS}$ characters is
\begin{align}
S_{(s_1, t_1),(s_2, t_2)} = \frac{2}{\sqrt{p p'}} \left( \cos(\frac{2\pi \lambda_1 \lambda_2}{4 p p'})-\cos(\frac{2\pi \bar{\lambda}_1 \lambda_2}{4 p p'})  \right)\;, \label{S-matrix}
\end{align}
with $\lambda_i = p t_i - p' s_i,\bar{\lambda}_i  = p t_i + p's_i$. The S-matrix determines the transformation rule of the NS characters under the modular $S$-transformation, $\tau\rightarrow -\frac{1}{\tau}$:
\begin{align}
\chi_\alpha (\tilde{q}) =\sum_\beta S_{\a\b} \chi_\beta (q)\;\;\textrm{ where } q:=e^{2\pi i \tau} \textrm{ and }\tilde{q} :=e^{2\pi i (-1/\tau)} \label{S-matrix : modular}\;.
\end{align}
Here $\alpha$ and $\beta$ label the NS primaries.  
\paragraph{3D Rank-0 $\CN=4$ SCFT $\CT_{(p,p')}$} Let the $\CT_{(p,p')}$ be the bulk 3D rank-0 SCFT associated to the minimal model $SM(p,p')$  via the non-unitary bulk-boundary correspondence:
\begin{align}
\textrm{3D SCFT } \CT_{(p,p')}\xrightarrow{\rm top'l\; A-twisting\;} \textrm{3D TQFT } \CT^{\rm A}_{(p,p')}\xrightarrow{\textrm {at boundary}} \textrm{2D RCFT } SM(p,p'). \label{CT(p,p')}
\end{align}
Basic dictionaries of the correspondence are summarized in the Table \ref{Table : Dictionaries}. Here $\mathbb{B}$ is a supersymmetric boundary condition of the rank-0 SCFT which becomes a holomorphic boundary condition in the A-twisted theory $\CT^{\rm A}$, under which the non-unitary TQFT $\CT^{\rm A}$ supports (chiral) rational conformal field theory $\CR[\CT^{\rm A};\mathbb{B}]$ at the boundary.  Refer to \cite{Gang:2021hrd,Gang:2023rei,Gang:2023ggt} for details and reasons behind the dictionaries.  Since we are interested in the case when the boundary RCFT is a supersymmetric minimal model, we list the dictionaries for fermionic case. One crucial difference from the bosonic case is that the conformal dimensions $h_\alpha$ of primaries  are determined modulo $1/2$, instead of $1$, from the bulk computation. 
\begin{table}[ht]
	\begin{center}
		\begin{tabular}{|c|c|}
			\hline
			$\quad \textrm{Boundary 2D $\chi$RCFT } \mathcal{R}[\CT^A, \mathbb{B}]$ \quad &  \textrm{Bulk 3D rank-0 SCFT }$\CT$ 
			\\
			\hline
			\hline
			& \textrm{Bethe vacua } $\{\vec{z}_\alpha\}_{\a=0}^{N-1}$
			\\
			NS-sector primaries $\CO_{\alpha =0,\ldots, N-1}$  & \textrm{ or }
			\\
			& \textrm{BPS loop operators $\{ L_\alpha (\vec{z}) \}_{\alpha=0}^{N-1}$}
			\\
			\hline
			$(S^{}_{0\alpha})^{-2}$ &    $ \CH(\vec{z}_\alpha, M=0, \nu=-1)$
			\\
			\hline
			$|S^{}_{0 0}|$ &   $\bigl|\CZ_{S^3_b}(M=0, \nu=-1)\bigr|$ 
			\\
			\hline
				${\rm min}_\alpha |S^{}_{0 \alpha}|$ &   $e^{-F}:=\bigl|\CZ_{S^3_b}(M=0, \nu=0)\bigr|$ 
			\\
			\hline
			$W_\beta(\alpha) :=S^{}_{\alpha \beta}/S^{}_{0 \beta}$ & $  L_\alpha  (\vec{z}_\beta)$ 
			\\
			\hline
			Conformal dimension $h_{\alpha}$&  $\left(\frac{\CF(\vec{z}_\alpha, M=0, \nu=-1)}{\CF(\vec{z}_{\alpha=0}, M=0, \nu=-1)} \right)^2 =\exp (4\pi i h_\alpha)$
			\\
			\hline
			$\chi^{}_\alpha (q) = q^{h_\alpha - c/24} I^{L_\alpha}_{\rm half:\mathbb{B}}$  &    Half-index $I_{\rm half:\mathbb{B}}^{L_\alpha} (q, \eta=1, \nu=-1)$
			\\
			\hline
		\end{tabular}
	\end{center}
	\caption{Basic dictionaries of non-unitary bulk boundary  correspondence for fermionic case, i.e. the boundary theory  is a  fermionic (or spin) RCFT. Here $\{\chi_\alpha\}_{\alpha=0}^{N-1}$ are the conformal characters, $\textrm{Tr} q^{L_0 - \frac{c}{24}}$, of  irreducible modules associated to   primaries $\{\CO_\alpha\}$  and $\alpha=0$ corresponds to the vacuum module, i.e. $\CO_{\alpha=0}$ is the identity operator. Supersymmetric quantities ($\CH,\CF,\CZ_{S^3_b}$ and $I^L_{\rm half} $) of the bulk rank-0 SCFT appearing in the table are explained in  main text.  $F$ is the round 3-sphere free energy \cite{Jafferis:2011zi,Casini:2015woa}.}
	\label{Table : Dictionaries}
\end{table}

\paragraph{BPS partition functions of rank-0 SCFTs} Now let us explain the supersymmetric quantities appearing in the right-hand side of the table. They can be computed  using the so-called supersymmetric localization method, which is applicable to any 3D $\CN=2$ theories. 
In terms of an $\CN=2$ supersymmetry subalgebra, rank-0 SCFT has a $U(1)_A$ flavor symmetry whose charge $A$ is 
\begin{align}
A = (J_3^C- J_3^H)\;,
\end{align}
where $J_3^C$ and $J_3^H$ are the two Cartans of the $SO(4) \simeq SU(2)^C\times SU(2)^H$ R-symmetry normalized as $J_3 \in \mathbb{Z}/2$.
Supersymmetric partition functions of rank-0 SCFT depends on $M$ (or $\eta$) and $\nu$ where $M$ (or $\eta$) is a (rescaled) real mass parameter  (or fugacity) for the $U(1)_A$ symmetry and $\nu$ parametrizes the R-symmetry mixing as follows
\begin{align}
R_\nu = (J_3^C+J_3^H) + \nu (J_3^C-J_3^H)\;. \label{R and A from SO(4)}
\end{align}
 $R_{\nu=0}$ corresponds to the superconformal R-charge.  We consider 3 types of supersymmetric backgrounds on closed 3-manifolds, superconformal index \cite{Kim:2009wb,Imamura:2011su}, squashed 3-sphere partition function \cite{Kapustin:2009kz,Jafferis:2010un,Hama:2010av,Hama:2011ea} and twisted partition functions \cite{Benini:2015noa,Benini:2016hjo,Closset:2016arn,Closset:2017zgf,Closset:2018ghr}. We also consider half-index \cite{Gadde:2013wq,Gadde:2013sca,Yoshida:2014ssa,Dimofte:2017tpi} defined on $D_2\times S^1$ with a proper SUSY boundary condition $\mathbb{B}$. 
 
 The superconformal index $\CI_{\rm sci}(q; \eta, \nu)$ is defined as
\begin{align}
\CI_{\rm sci} (q; \eta, \nu) := \textrm{Tr}_{\CH_{\rm rad}(S^2)} (-1)^{R_\nu} q^{\frac{R_\nu}2 +j_3} \eta^{A}\;,
\end{align}
where the trace is taken over the radially quantized Hilbert-space $\CH_{\rm rad}(S^2)$, whose elements are in one-to-one with local operators of the SCFT. $j_3$ is the 3rd component of the Lorentz spin.  Due to the $(-1)^{R_{\nu}}$ factor,  there are huge cancellations and only local  operators  satisfying following condition could give non-vanishing contributions to the index,
\begin{align}
\Delta = R_{\nu=0}+j_3\;, \label{BPS condition for index}
\end{align}
where $\Delta $ is the conformal dimension.  The squashed 3-sphere partition $\CZ_{S^3_b}(b^2; M, \nu)$ is defined on the following $S^3_b$ background:
\begin{align}
S^3_b  := \{(z,w)\in \mathbb{C}^2 \;:\; b^2|z|^2+b^{-2}|w|^2=1\}\;.
\end{align}
When $b=1$, it corresponds to the round 3-sphere.
The partition function depends on the rescaled real mass parameter $M$ ($b\times $real mass) and R-symmetry mixing parameter $\nu$ only through a holomorphic combination $M+(i \pi +\frac{\hbar}2)\nu $:
\begin{align}
\mathcal{Z}_{S^3_b}(b^2;M, \nu)  = \mathcal{Z}_{S^3_b}\left(b^2;M+ (i \pi +\frac{\hbar}2) \nu\right)\;, \label{holomorphy}  
\end{align}
 where  $\hbar:=2\pi i b^2$.
The round 3-sphere partition function can be used in determining the correct IR superconformal R-charge via F-maximization \cite{Jafferis:2010un} and in defining the $F =-\log |\CZ_{S^3_b}(b^2=1)|$, which is a proper measure of the number of degrees of freedom  in  3D CFTs \cite{Jafferis:2011zi,Casini:2015woa}.  

The twisted partition function $\CZ_{\CM_{g,p}} (M, \nu)$ on $\CM_{g,p}$, degree $p$ circle bundle over genus $g$ Riemann surface $\Sigma_g$, can be  given in the following form
\begin{align}
\begin{split}
&\CZ_{\CM_{g,p\in 2\mathbb{Z}}} (M, \nu) = \sum_{\vec{z}_\alpha \in \CS_{\rm BE}(M, \nu)\;} \left(\CH (\vec{z}_\alpha ; M, \nu)\right)^{g-1} \left(\CF(\vec{z}_\alpha ; M, \nu)\right)^p\;,
\\
&\CS_{\rm BE} (M, \nu):=  \{\vec{z}\;:\; \vec{P} (\vec{z}; M , \nu)=1 \textrm{ and }\; w(\vec{z}) \neq \vec{z} \textrm{ for all $\omega \in \textrm{W}(G)$}\}/\textrm{W(G)}\;.
\end{split}
\end{align}
Here $\vec{P}(\vec{z};M, \nu) = 1$ is a set of  algebraic equations on $\vec{z}=(z_1, \ldots, z_r)$ called Bethe equations, whose solution $\vec{z}_\alpha$ is called Bethe-vacuum. Both of the size $r$ of the vector $\vec{z}$ and $\vec{P}$ are equal to the rank of gauge group $G$. $W(G)$ denotes the Weyl subgroup of $G$, which acts on $\vec{z}$. 
There are two distinct SUSY backgrounds on $\CM_{g, p \in 2\mathbb{Z}}$ depending on the spin-structure choices along the fiber $[S^1]$-direction \cite{Closset:2018ghr}. $(\CH,\CF)$ are so-called (handle gluing, fibering) operators and they depend on the spin-structure choice. In the dictionary,  we use $(\CH,\CF)$ in  anti-periodic boundary condition, which corresponds to $\nu_R = \frac{1}2 \; (\textrm{mod 1})$ in  \cite{Closset:2018ghr}.  One should  use the $(\CH, \CF)$ in the periodic boundary condition, i.e. $\nu_R = 0 \; (\textrm{mod 1})$ to compute the twisted partition function with odd $p$. 
When $p=0$, the $\CM_{g,p}= \Sigma_{g}\times S^1$ and the twisted partition function becomes  twisted index $\CI_{\Sigma_g}$ on $\Sigma_g$:
\begin{align}
\CI^{\rm }_{\Sigma_g} (\eta= e^M , \nu)= \CZ_{\CM_{g,p=0}} (M, \nu) = \textrm{Tr}_{\CH (\Sigma_g ; R_\nu)} (-1)^{R_\nu} \eta^{A}\;.
\end{align}
Here $\CH(\Sigma_g;R_\nu)$ is the Hilbert-space on $\Sigma_g$ with background monopole flux coupled to the R-symmetry is turned on as follows 
\begin{align}
\int_{\Sigma_g} F_{R_{\nu}} = (2 -2g)\;,
\end{align}
to preserve some supercharges. Due to the Dirac quantization condition for the $U(1)$ R-symmetry,  the $\nu$ can take only following discrete values in the twisted index:
\begin{align}
\CI_g (\eta, \nu) \textrm{ is well-defined only when } (2-2g) R_\nu \in \mathbb{Z}\;.
\end{align}
For $\CN=4$ SCFTs, the R-charge $R_{\nu = \pm 1}$ in the A/B-twisting limits always satisfies the quantization condition for all $g$ since $R_{\nu =\pm 1} = 2J_3^{C/H} \in \mathbb{Z}$. Thus, the handle-gluing operators are also well-defined in the limits and the dictionary  for them in Table \ref{Table : Dictionaries} makes sense. 

 Half-index $I_{\rm half:\mathbb{B}} (q; \eta, \nu)$ with a supersymmetric boundary condition $\mathbb{B}$ is defined as\footnote{The boundary condition $\mathbb{B}$ could preserve a subgroup $G_{\mathbb{B}}$ of gauge group $G$. In the case, one can introduce fugacities and R-symmetry mixing parameters for the unbroken gauge group in the half-index. The boundary chiral RCFT $\chi{\rm RCFT}[\CT^{A};\mathbb{B}]$ has the subgroup $G_{\mathbb{B}}$ as a flavor symmetry. To realize the minimal models $SM(p,p')$, we consider a boundary condition $\mathbb{B}$ such that the unbroken gauge group $G_{\mathbb{B}}$ is at most a finite discrete group. }
\begin{align}
I_{\rm half :\mathbb{B}} (q; \eta, \nu) := \textrm{Tr}_{\CH(HS^2;\mathbb{B})} (-1)^{R_\nu} q^{\frac{R_\nu}2 + j_3} \eta^A\;.
\end{align}
Here $\CH(HS^2;\mathbb{B})$ is the Hilbert-space on the (northern) hemisphere $HS^2$, which is  topologically  a disk, with a supersymmetric boundary condition $\mathbb{B}$ on the boundary $\partial (HS^2)$. The half-index can be decorated by inserting temporal supersymmetric loop operator $L$ at the north pole:  
\begin{align}
I^{L}_{\rm half :\mathbb{B}} (q; \eta, \nu) := \textrm{Tr}_{\CH(HS^2+L;\mathbb{B})} (-1)^{R_\nu} q^{\frac{R_\nu}2 + j_3} \eta^A\;.
\end{align}
Here $\CH(HS^2+L; \mathbb{B})$ is the Hilbert-space on $HS^2$ with an insertion of the loop operator $L$. 

For rank-0 SCFTs, the supersymmetric partition functions in the limit,  $M\rightarrow 0$ (or $\eta \rightarrow 1$) and $\nu \rightarrow  -1$ (resp. $\nu \rightarrow  +1$), are known to reproduce the partition functions of the A-twisted (resp. B-twisted) of the theory $\CT^{\rm A}$ (resp. $\CT^{\rm B}$). We call the two limits A/B-twisting limits:
\begin{align}
\begin{split}
&\textrm{A-twisting limit : } M\rightarrow 0 \;(\textrm{or }\eta \rightarrow 1) \textrm{ and }\nu \rightarrow -1\;, 
\\
&\textrm{B-twisting limit : } M\rightarrow 0 \;(\textrm{or }\eta \rightarrow 1) \textrm{ and }\nu \rightarrow +1\;.
\end{split}
\end{align}

In the rest of the section, we propose UV field theory for $\CT_{(p,p')}$ with $p=2$ and $p=3$ and test the proposal by checking the dictionaries listed in the Table \ref{Table : Dictionaries}. 

\subsection{Bulk dual rank-0 SCFT of $SM(2, 4r)$}
We propose three distinct UV gauge theory descriptions of the rank-0 SCFT $\CT_{(2,4r)}$, which are related to each other by IR dualities modulo a decoupled invertible spin TQFT $SO(1)_1$ (also known as Ising spin-TQFT), whose boundary RCFT is a free Majornara-Weyl fermion theory (a.k.a fermionized Ising CFT).  More precisely, we propose that
\begin{align}
\begin{split}
&\left( \widetilde{\CT}_{(2,4r)}/\widetilde{\CT}'_{(2,4r)}  \textrm{ in \eqref{UV theory for (2,4r)-I}/ \eqref{UV theory for (2,4r)-III}}\right) \xrightarrow{\;\textrm{A-twisting/bulk-boundary}\;} SM(2,4r) \otimes (\textrm{Free  fermion}),
\\
&\left( \CT_{(2,4r)}   \textrm{ in \eqref{UV theory for (2,4r)-II}} \right) \xrightarrow{\;\textrm{A-twisting/bulk-boundary}\;} SM(2,4r)\;. \label{proposal for SM(2,4r)}
\end{split}
\end{align}
and 
\begin{align}
\begin{split}
&\left( \widetilde{\CT}_{(2,4r)}  \textrm{ in \eqref{UV theory for (2,4r)-I} }\right) \simeq \left( \widetilde{\CT}'_{(2,4r)}  \textrm{ in   \eqref{UV theory for (2,4r)-III}}\right) \simeq \left( \CT_{(2,4r)}  \textrm{ in   \eqref{UV theory for (2,4r)-II}}\right) \otimes SO(1)_1\;, \label{dualities among SM(2,4r)}
\end{split}
\end{align}
where $\simeq $ means the IR equivalence.  The tilde in $\widetilde{\CT}$ is to distinguish it from $\CT_{(p,p')}$ in \eqref{CT(p,p')}. Two theories are related as $\widetilde{\CT} \simeq \CT \otimes SO(1)_1$. 
The decoupled $SO(1)_1$ is almost invisible in the bulk BPS partition functions since its partition functions on 3-manifolds are purely phases and bulk BPS partition functions  have overall phase factor ambiguities.  The decoupled sector can be detected from the half-index computation, to which the invertible spin-TQFT contributes by an overall factor  $\chi_F (q)$, the character of free Majorana-Weyl fermion theory, given in \eqref{free-fermion character}. The free  fermion theory has only unique NS primary, identity operator, and trivial modular  NS-NS $S$-matrix, i.e. $S=1$, and central charge $c=\frac{1}2$.
Interestingly,  
the  theories  have actually $\CN=5$ superconformal symmetry in the IR.  

\subsubsection{UV description I  }
The first UV gauge theory description is ($r\geq 2$)\footnote{When $r=1$, $K=(1)$ with the monopole superpotential $\CW=V_{\mathbf{m}=(2)}$, the gauge theory has a mass gap and flows to the Ising spin-TQFT, $SO(1)_1$, in the IR.  }
\begin{align}
\begin{split}
&\widetilde{\CT}_{(2,4r) }:= \frac{(\CT_\Delta)^{\otimes r}}{[U(1)^{r}_{\mathbf Q}]_{K} }\textrm{ with superpotential } \CW = \CO_{(\mathbf{m}_1, \mathbf{n}_1)}+\ldots +\CO_{(\mathbf{m}_{r-1}, \mathbf{n}_{r-1})}\;,
\\
&\textrm{where}
\\
& K= \begin{pmatrix}
	1&-1&-1&\cdots&-1&-1 \\
	-1&2&2&\cdots&2&2 \\
	-1&2&4&\cdots&4&4 \\
	\vdots&\vdots&\vdots&\ddots&\vdots&\vdots\\
	-1&2&4&\cdots&2(r-2)&2(r-2)\\
	-1&2&4&\cdots&2(r-2)&2(r-1)
\end{pmatrix}, \quad \mathbf{Q} = \mathbb{I}_{r\times r},
\\
&\mathbf{m}_1 = (2,\mathbf{0}_{r-1}), \; \mathbf{m}_2 = (0,2,-1,\mathbf{0}_{r-3}), \; \mathbf{m}_{3 \leq I \leq r-1} = (\mathbf{0}_{I-2},-1,2,-1,\mathbf{0}_{r-I-1}) \; ,
\\
&\mathbf{n}_1 = (0,\mathbf{2}_{r-1}), \; \mathbf{n}_2 = (1,\mathbf{0}_{r-1}), \; \mathbf{n}_{3 \leq I \leq r-1} = (\mathbf{0}_{r}) \; .
\end{split} \label{UV theory for (2,4r)-I}
\end{align}
%
 The theory $\CT_{\Delta}$ is a  free  theory of single 3D $\CN=2$ chiral multiplet $\Phi$  with  Chern-Simon (CS) level $-\frac{1}2$ \cite{Dimofte:2011ju} for the background gauge field coupled to the $U(1)$ flavor symmetry. 
 Its Lagrangian using superfields is given as
 \begin{align}
 \mathcal{L}_{\Delta} (\Phi;V) =  \int d^4 \theta \left(\Phi^\dagger e^{V} \Phi- \frac{1}{8\pi }  \Sigma_V V \right)\;. \label{T-Delta}
 \end{align}
 Here $\Sigma_V$ is the field strength multiplet of the background vector multiplet $V$ coupled to the $U(1)$ flavor symmetry.  The first term gives the kinetic terms for the chiral field coupled to the background $U(1)$ gauge field and the 2nd gives the background CS term.  The theory also has a background mixed CS level $1/2$ (which is not written in the above Lagrangian) between $U(1)$ flavor symmetry and $U(1)$ R-symmetry when the R-symmetry is chosen to be $R(\Phi)=0$. 
 
 The $(\CT_\Delta)^{\otimes r}$, $r$-copies of $\CT_\Delta$, has $U(1)^r$ flavor symmetry and $/U(1)^r_\mathbf{Q}$ denotes the $\CN=2$ gauging of  the $U(1)^r$ flavor symmetry  with mixed  CS  level $K$ and charge matrix $\mathbf{Q}$:
\begin{align}
\mathbf{Q}_{ab} = (\textrm{Charge of $b$-th chiral multiplet $\Phi_b$ under $a$-th $U(1)$ gauge symmetry})\;.
\end{align} 
Throughout this paper, we consider the charge matrix $\mathbf{Q}$  given by a diagonal matrix
\begin{align}
\mathbf{Q} = \textrm{diag}\{Q_1, \ldots, Q_r\}\;.
\end{align}
Taking into account of the background CS level $-1/2$ in the $\CT_\Delta$  theory, the gauge theory is nothing but
\begin{align}
\begin{split}
&\frac{(\CT_\Delta)^{\otimes r}}{[U(1)^{r}_{\mathbf Q}]_{K} }=\; U(1)^r\; \textrm{gauge theory coupled to $r$ chiral multiplets of charge $\mathbf{Q}$ with} 
\\
& \quad \quad \qquad \qquad \textrm{mixed CS level } K - \frac{1}2  \mathbf{Q} \mathbf{Q}^T \;,\;\; 
\end{split}
\end{align}
whose Lagrangian is
\begin{align}
\CL (\vec{W}) = \sum_{a=1}^r \CL_\Delta (\Phi_a;  Q_a v_a) +\int d^4 \theta \left( \sum_{a,b=1}^r \frac{K_{ab}}{4\pi} \Sigma_{v_a} v_b +\sum_{a=1}^r \frac{1}{2\pi} \Sigma_{v_a}W_a  \right)\;.
\end{align}
Here $\{v_a\}_{a=1}^r$ are  dynamical vector mutiplets for $U(1)^r$ gauge symmetry while $\{W_a\}_{a=1}^r$ are background vector multiplets coupled to  $U(1)^r$ topological symmetry. 

$\CO_{(\mathbf{n},\mathbf{m})}$  with $\mathbf{m}=(m_1, m_2,\ldots, m_r) \in \mathbb{Z}^r$ and $\mathbf{n}=(n_1, n_2,\ldots, n_r) \in \mathbb{Z}_{\geq 0}^r$  denotes a dressed  BPS monopole operator of the form
\begin{align}
\CO_{(\mathbf{m},\mathbf{n})} = \left(\prod_{a}(\Phi_a)^{n_a} \right) V_{\mathbf{m}}\;,
\end{align}
where the $V_{\mathbf{m}}$ is half-BPS bare monopole operator with magnetic flux $\mathbf{m}$.
The  charge $q_a$ of the monopole operator under the $a$-th $U(1)$ gauge group  is (see, for example, \cite{Benini:2011cma})
\begin{align}
q_a (\CO_{(\mathbf{m},\mathbf{n})})= (\mathbf{Q} \cdot \mathbf{n})_a + \left( (K- \frac{1}2 \mathbf{Q} \mathbf{Q}^T ) \cdot \mathbf{m}  \right)_a -  \frac{1}2 \sum_b \mathbf{Q}_{ab} |(\mathbf{Q}^T\cdot \mathbf{m})_b|\;.
\end{align}
A gauge invariant BPS monopole operator $\CO_{(\mathbf{m},\mathbf{n})}$  is 1/2 BPS chiral primary operator if
\begin{align}
n_a m_a=0, \quad \textrm{for all }a=1,\ldots, r\;,
\end{align}
i.e. purely magnetic or purely electric under all the $U(1)$ factors in the gauge symmetry.  Notice that the monopole operators appearing in the superpotential are all gauge-invariant  1/2 BPS chiral primary operators. 

Before the superpotential deformation, the theory $ \frac{(\CT_\Delta)^{\otimes r}}{[U(1)^{r}_{\mathbf Q}]_{K} }$ has $U(1)^r$ topological symmetry, whose charges $\{T_{a}\}_{a=1,\ldots, r}$ are the monopole charges of $U(1)^r$ gauge symmetry.   After the monopole superpotential deformations,   the $\CN=2$ gauge theory has only a $U(1)$ flavor symmetry, $U(1)_A$, whose charge $A$ is given as
\begin{align}
A = \vec{A}\cdot \vec{T}=   \sum_{a=1}^r (a-1)T_a  =  T_2 + 2T_3+\ldots +(r-1)T_r \;.
\end{align}
One can check that $\vec{A}\cdot \mathbf{m}_I =0$ for all $I=1,\ldots, r-1$. 

The UV gauge theory has only manifest $\CN=2$ supersymmetry and we will claim that the  theory has emergent supersymmetries in the IR and flows to a 3D rank-0 $\CN=4$ (actually $\CN=5$) SCFT. In the SUSY enhancement, the $U(1)_R \times U(1)_A$ symmetry is  enhanced to $SO(5)$ R-symmetry. The  two UV $U(1)$s are embedded into the $SO(4)\subset SO(5)$ R-symmetry as in \eqref{R and A from SO(4)} and embedded into the $SO(5)$ R-symmetry as follows
\begin{align}
\textrm{$R_{\nu=0}$ and $A$ are Cartans of $SO(2)$ and $SO(3)$ of $SO(2)\times SO(3) \subset SO(5)$ respectively.} \label{R and A in SO(5)}
\end{align}

\begin{figure}[h]
	\centering
	\includegraphics[width=0.495\textwidth]{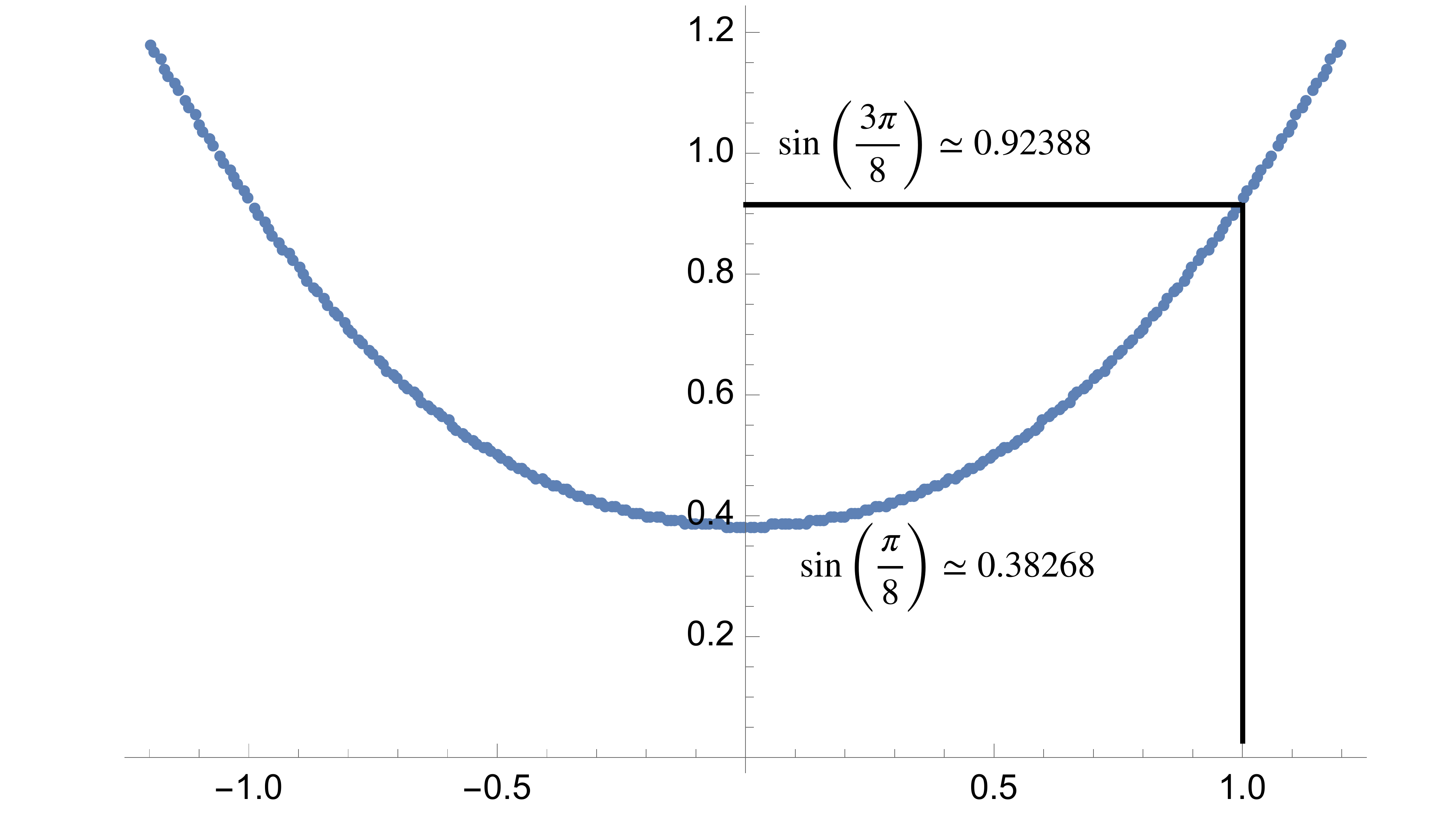}
	\includegraphics[width=0.495\textwidth]{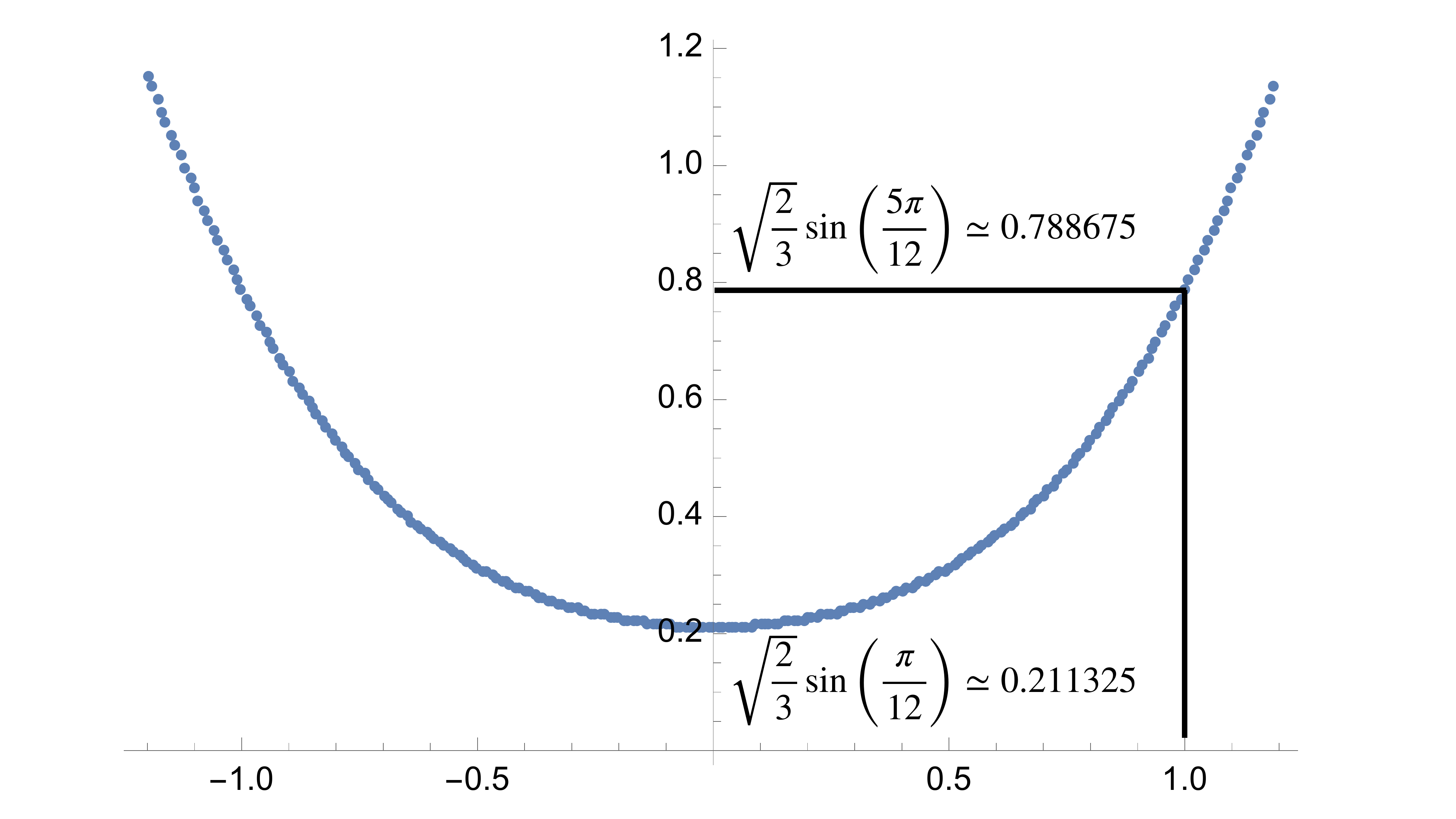}
	\caption{Graph of $|Z_{S^3_b} (b^2=1, M=0, \nu)|$ for $\widetilde{\CT}_{(2,8)}$ theory (Left) and for $\widetilde{\CT}_{(2,12)}$ theory (Right).  The $|Z_{S^3_b} (b^2=1, M=0, \nu)|$s are  even functions in $\nu$ and minimized at $\nu=0$. The values of  $|Z_{S^3_b} (b^2=1, M=0, \nu=\pm 1)|$ are identical to the $|S_{\alpha=0, \beta=0} = S_{(1,1),(1,1)}|$ of $SM(2,8)$ and $SM(2,12)$ respectively.   The value of  $|Z_{S^3_b} (b^2=1, M=0, \nu=0)|$ is identical to the $\textrm{min}_\beta|S_{\alpha=0, \beta} |$ of $SM(2,8)$ and $SM(2,12)$ respectively. We draw the graph using the Bethe-sum formula in   \eqref{Bethe-sum of ZS3} combined with \eqref{Bethe-sum T(2,4r)}.  }
	\label{fig:F-maximization1}
\end{figure}

\paragraph{Squashed 3-sphere partition function} The partition function of the theory before the superpotential deformation, i.e. $(\CT_\Delta)^{\otimes r}/[U(1)^r_\mathbf{Q}]_K$, is $(\hbar := 2\pi i b^2)$
\begin{align}
\begin{split}
\mathcal{Z}^{(K,Q)}_{S^3_b} (\vec{M}, \vec{\nu}) &= \int \frac{d^r \vec{Z}}{(2\pi \hbar)^{r/2}} \prod_{a=1}^r \psi_\hbar (Q_a Z_a)\exp \left( \frac{\vec{Z}\cdot K  \cdot \vec{Z}+ 2 \vec{Z}\cdot \vec{W} }{2\hbar}\right)\bigg{|}_{\vec{W} = \vec{M} +(i \pi +\frac{\hbar}2) \vec{\nu}}\;,
\\
& = \int \frac{d^r \vec{Z}}{(2\pi \hbar)^{r/2}} \CI_\hbar^{(K,Q)}  (\vec{Z};\vec{M},\vec{\nu})\;.\label{S^3_b (K,Q)}
\end{split}
\end{align}
The special function $\psi_\hbar(Z)$ is the quantum dilogarithm (Q.D.L). It computes the squashed 3-sphere partition function of the $\CT_\Delta$ theory with $Z = M+(i \pi+\frac{\hbar}2 R(\Phi))$ where $M$ is the rescaled real mass for the $U(1)$ flavor symmetry. Its definition and basic  properties are reviewed in Appendix \ref{appendix}.
$\vec{M}$ is the (rescaled) real masses, i.e. FI parameters,  coupled to the $U(1)^r$ topological symmetry. $\vec{\nu}$ parameterize the mixing between R-symmetry and $U(1)^r$ topological symmetry.  The R-charge $R_{\vec{\nu}}$ at the mixing parameter is
\begin{align}
R_{\vec{\nu}} = R_* + \vec{\nu} \cdot \vec{T}\;.
\end{align}
Here $R_{*}$ is a reference R-charge. In the above expression, the reference R-charge is chosen as\footnote{The reference R-charge $R_*$ is chosen such that  $R_*(\Phi_a) =0$  for all $a=1,\ldots, r$ and the mixed CS level $(k_{gR})_a$ between $U(1)_R$ and $a$-th $U(1)$  gauge group is $\frac{1}2 Q_a$. Generally,  $R(\CO_{\mathbf{n}, \mathbf{m}}) =\sum_a \left( n_a R(\Phi_a)+ \frac{1-R(\Phi_a)}2  |Q_a m_a|+(k_{gR})_a m_a \right)$ \cite{Benini:2011cma}. }
\begin{align}
&R_{*} (\CO_{(\mathbf{n}, \mathbf{m})}) = \frac{1}2 \sum_{a=1}^r  (Q_a m_a +|Q_a m_a|)\;.
\end{align}
After the deformation with the monopole superpotentials, one need to impose following conditions on $(\vec{M}, \vec{\nu})$:
\begin{align}
\begin{split}
&\vec{M}\cdot \mathbf{m}_I =0\;,
\\
&
R_{\vec{\nu}} (\CO_{(\mathbf{m}_I, \mathbf{n}_I)}) = \frac{1}2 \sum_{a=1}^r (Q_a (m_I)_a +|Q_a (m_I)_a|) + \vec{\nu} \cdot \mathbf{m}_I = 2\;, \label{constraints by superpotential}
\end{split}
\end{align}
for all $I=1,\ldots, r-1$. Solving the equations, the $(\vec{M}, \vec{\nu})$ can be parameterized as
\begin{align}
\vec{M} = \vec{A} M\;, \quad \vec{\nu} = (\nu-1) \vec{A}\;, \label{(M, nu)-parametrization}
\end{align}
and the  partition function for the $\widetilde{\CT}_{(2,2r)}$ theory becomes
\begin{align}
\begin{split}
\mathcal{Z}_{S^3_b} (b^2,M,\nu) &= \mathcal{Z}^{(K,Q)}_{S^3_b} \left(\vec{M} = \vec{A} M,\vec{\nu} = (\nu-1) \vec{A}\right)\;,
\\
& = \int \frac{d^r \vec{Z}}{(2\pi \hbar)^{r/2}} \CI_\hbar (\vec{Z}; M, \nu)\;. \label{squashed 3-sphere ptn T(2,4r)}
\end{split}
\end{align}
$(M,\nu)$ is the (real mass, R-symmetry mixing parameter) for the $U(1)_A$ symmetry.
\begin{align}
R_\nu = R_{\nu=0}+ \nu A = R_* + (\nu-1) \vec{A} \cdot \vec{T}\;.
\end{align}
 Superconformal R-charge, $R_{\nu= \nu_{\rm IR}}$, of the IR fixed point can be  determined using F-maximization \cite{Jafferis:2010un}. Namely, 
\begin{align}
\label{eq:fmax}
F := -\log |\mathcal{Z}_{S^3_b}(b^2=1,M=0, \nu)| \textrm{ is maximized at $\nu= \nu_{\rm IR}$}.
\end{align}
In our case, we choose the parameterization as in \eqref{(M, nu)-parametrization}  in a way that the $\nu=0$ corresponds to the superconformal R charge, $\nu_{\rm IR} = 0$, see Figure \ref{fig:F-maximization1}. The squashed 3-sphere partition function has an  overall phase factor ambiguity of the following form
\begin{align}
\exp \left( i \pi (b^2+b^{-2}) \delta_1+i \pi\delta_2\right)  \textrm{ with }\delta_1, \delta_2 \in \mathbb{Q}\;,\label{phase factor ambiguity}
\end{align}
which depends on the choice of background CS level of R-symmetry, 3-manifold framming choice and etc.  

\paragraph{Superconformal index} The superconformal index for the theory before the monopole superpotential deformation is 
\begin{align}
\begin{split}
&\CI_{\rm sci}^{(K,Q)} (q;\vec{\eta},\vec{\nu}) 
\\
&= \sum_{\mathbf{m} \in \mathbb{Z}^r} \left( \prod_{a=1}^r \oint \frac{du_a}{2\p i u_a}  \right)  \left( \prod_{a=1}^r  \CI_\Delta (Q_a m_a, u_a^{Q_a}) (\eta_a (-q^{1/2})^{\nu_a})^{m_a} \right) \left( \prod_{a,b=1}^r u_a^{K_{ab} m_b}\right)\;. \label{SCI (K,Q)}
\end{split}
\end{align}
Here $\CI_\Delta (m,u)$ is the tetrahedron index introduced in \cite{Dimofte:2011py}, which computes the generalized superconformal index \cite{Kapustin:2011jm} of the $\CT_\Delta$ theory. See appendix \ref{appendix} for details.
After the monopole operator superpotential deformation, the superconformal index for the $\widetilde{\CT}_{(2,4r)}$ theory is
\begin{align}
\CI_{\rm sci} (q;\eta, \nu) = \CI^{(K,Q)}_{\rm sci} \left(q;\vec{\eta}, \vec{\nu} = (\nu-1) \vec{A}\right)|_{\eta_a \rightarrow \eta^{A_a}}\;. \label{SCO T(2,4r)-1}
\end{align}
Using the expression above, one can compute the  index and find that
\begin{align}
\begin{split}
&r=2:
\\
&1-q^{1/2}-\left(1+\eta +\frac{1}{\eta }\right) q-\left(2+\eta +\frac{1}{\eta }\right) q^{3/2}-\left(2+\eta +\frac{1}{\eta }\right) q^2-q^{5/2}
\\
&+\left(\eta ^2+\frac{1}{\eta ^2}+\eta +\frac{1}{\eta }+1\right) q^3+\left(\eta ^2+\frac{1}{\eta ^2}+2 \eta +\frac{2}{\eta }+2\right) q^{7/2}+\ldots
\\
&r=3:
\\
&1-q^{1/2}-\left(1+\eta +\frac{1}{\eta }\right) q-\left(2+\eta +\frac{1}{\eta }\right) q^{3/2}-q^2+\left(\eta ^2+\frac{1}{\eta ^2}+2 \eta +\frac{2}{\eta }+2\right) q^{5/2}
\\
&+\left(2 \eta ^2+\frac{2}{\eta ^2}+4 \eta +\frac{4}{\eta }+6\right) q^3+\left(2 \eta ^2+\frac{2}{\eta ^2}+6 \eta +\frac{6}{\eta }+8\right) q^{7/2}+\ldots
\\
&\textrm{for all $r\geq 2$}:
\\
&\CI_{\rm sci} (q;\eta, \nu=0) = 1-q^{1/2} - \left(1+\eta +\frac{1}\eta\right)q-\left(2+\eta +\frac{1}\eta\right)q^{3/2}+\ldots\;,
\end{split}
\end{align}
The index computation gives non-trivial evidences for the SUSY enhancement, $\CN=2 \rightarrow \CN=5$, in the IR.  First, only $q^{\mathbb{Z}/4}$-terms (actually only $q^{\mathbb{Z}/2}$) appear in the index which is compatible with the fact that $\frac{R_{\nu=0}}2 + j_3\in \frac{\mathbb{Z}}4 $ for any $\CN\geq 3 $ theory. This is quite non-trivial fact since the superconformal R-charge $R_{\nu=0}$ is  determined by extremizing the function $F(\nu) = \log |\CZ_{S^3_b} (b^2=1, M=0, \nu)|$, which is highly non-trivial function as drawn in Figure \ref{fig:F-maximization1}. Second, the index contains  contributions from $\CN=5$ stress-energy tensor multiplet. The $\CN=5$ multiplet can be decomposed into several superconformal multiplets of a $\CN=2$ subalgebra (which can be identified as the UV supersymmetry). They include \cite{Cordova:2016emh}
\begin{align}
\begin{split}
&a)\;\textrm{Chiral primary multiplet with $\Delta =1$  in $\mathbf{1}$ of $SO(3)$}\;,
\\
&b)\;\textrm{Conserved current multiplet  in $\mathbf{3}$ of $SO(3)$}\;,
\\
&c)\;\textrm{Extra-SUSY current multiplet   in $\mathbf{3}$ of $SO(3)$}\;,
\end{split}
\end{align}
where $SO(3)$ is the $SO(3)$ in $SO(2)\times SO(3) \subset SO(5)$ R-symmetry which is flavor symmetry in terms of the UV $\CN=2$ supersymmetry. The $\CN=2$ multiplets contribute to the index as follows 
\begin{align}
\begin{split}
&a) \Rightarrow -q^{1/2} - q^{3/2} +\ldots\;,
\\
&b) \Rightarrow -\left(1+\eta+\frac{1}\eta\right)q+\ldots\;,
\\
&c) \Rightarrow -\left(1+\eta+\frac{1}\eta\right)q^{3/2}+\ldots\;,
\end{split}
\end{align}
which can be obtained from the explicit multiplet structure in \cite{Cordova:2016emh}.
One can see that all these terms appear in the superconformal index. Even though the appearance do not guarantee the SUSY enhancement \cite{Evtikhiev:2017heo}, since other $\CN=2$ multiplets could give the same contributions, it provides non-trivial circumstantial evidence. As another non-trivial evidence,  we will propose dual description in section \ref{sec: UV III} which has manifest $\CN=5$ supersymmetry. For  all $r\geq 2$, the indices in the A- and B-twisting limits become trivial, i.e. 
\begin{align}
&\CI_{\rm sci} (\eta =1, \nu=\pm 1) = 1\;.
\end{align}
It implies that the IR SCFT is of rank 0, if the SUSY enhancement really occurs, since the index in the A/B twisting limits compute the Coulomb/Higgs branch Hilbert-series \cite{Razamat:2014pta}.

\paragraph{Twisted partition functions} To compute the twisted partition function, we first consider the integrand $\CI^{(K,Q)}_\hbar$ of the squashed 3-sphere partition function \eqref{S^3_b (K,Q)} in the limit of $\hbar \rightarrow 0$ using \eqref{asymptotic of QDL}:
\begin{align}
\begin{split}
&\log \CI^{(K,Q)}_{\hbar} (\vec{Z};\vec{M}, \vec{\nu}) \xrightarrow{\quad \hbar \rightarrow 0 \quad }  \frac{1}\hbar \CW_0^{(K,Q)} (\vec{Z};\vec{M},\vec{\nu}) +\CW_1^{(K,Q)} (\vec{Z};\vec{M},\vec{\nu}) +O(\hbar) 
\\
&\textrm{ with }
\\
& \CW^{(K,Q)}_0 =\sum_a {\rm Li}_2 (e^{-Q_a Z_a}) + \frac{1}{2} \vec{Z}\cdot K \cdot \vec{Z} + (\vec{M}+i \pi \vec{\nu})\cdot \vec{Z}\;,
\\
& \CW^{(K,Q)}_1  =- \frac{1}2\sum_a \log (1-e^{-Q_a Z_a})+ \frac{1}2 \vec{Z}\cdot \vec{\nu}\;.
\end{split}
\end{align}
Then, the twisted partition function can be computed as
\begin{align}
\mathcal{Z}^{(K,Q)}_{\CM_{g,p\in 2\mathbb{Z}}} (\vec{M}, \vec{\nu}) = \sum_{\vec{z} \in \CS_{\rm BE}(\vec{M},\vec{\nu})} \CH(\vec{z};\vec{M}, \vec{\nu})^{g-1} (\CF (\vec{z};\vec{M},\vec{\nu}))^p\;, \label{twisted-ptn-(K,Q)}
\end{align}
where the Bethe-vacua $\CS_{\rm BE}$ is the set of  solutions of following algebraic equations ($\partial_a := \partial_{Z_a}$)
\begin{align}
\begin{split}
&\CS_{\rm BE} (\vec{M}, \vec{\nu}) = \bigg{\{}\vec{z}=(z_1, \ldots, z_r) \;:\; P_a (\vec{z}; \vec{M}, \vec{\nu})=1 \textrm{  for all }a=1,\ldots, r\bigg{\}}\;,
\\
&\textrm{ where } P_a := \exp \left( \partial_{a}\mathcal{W}^{(K,Q)}_0\right)|_{\vec{Z}\rightarrow \log \vec{z}} = \left(  \prod_{b=1}^r z_b^{K_{ab}} \right) (1-z_a^{-Q_a})^{Q_a}e^{M_a+i\pi \nu_a} \;.
\end{split}
\end{align}
The handle-gluing and fibering operators are
\begin{align}
\begin{split}
&\CH (\vec{z};\vec{M},\vec{\nu}) = \left( \det_{a,b} \partial_a \partial_b \CW_0 \right) \exp \left(-2\CW_1 \right)\bigg{|}_{\vec{Z}\rightarrow \log \vec{z}}\;,
\\
&\CF (\vec{z};\vec{M},\vec{\nu})  = \exp \left( - \frac{\CW_0 -\vec{Z}\cdot \partial_{\vec{Z}}\CW_0- \vec{M} \cdot \partial_{\vec{M}}\CW_0  }{2\pi i }\right)\bigg{|}_{\vec{Z}\rightarrow \log \vec{z}}\;.
\end{split}
\end{align}
Then, twisted partition functions for the $\widetilde{\CT}_{(2,4r)}$ theory is given as
\begin{align}
\begin{split}
&\mathcal{Z}_{\CM_{g, p \in 2\mathbb{Z}}} (M, \nu)= \sum_{\vec{z}_\alpha \in \CS_{\rm BE}} \CH(\vec{z}_\alpha; M, \nu)^{g-1} \CF(\vec{z}_\alpha; M, \nu)^{p}\;,
\\
&\textrm{with }  \CH/\CF (\vec{z}_\alpha; M, \nu) := \CH/\CF(\vec{z}_\alpha; \vec{M}=\vec{A}M, \vec{\nu}= (\nu-1)\vec{A}) \;.
\end{split}
\end{align}
There are two distinct SUSY backgrounds on $\CM_{g, p \in 2\mathbb{Z}}$ depending on the spin-structure choices along the fiber $[S^1]$-direction \cite{Closset:2018ghr}. In the above computation, we  choose an anti-periodic, which corresponds to $\nu_R = \frac{1}2  (\textrm{mod 1})$ in  \cite{Closset:2018ghr}.  Due to the phase ambiguity of squashed 3-sphere partition function in \eqref{phase factor ambiguity}, the $\CF(\vec{z}_\alpha)$ and $\CH(\vec{z}_\alpha)$ have  $\alpha$-independent phase factor ambiguities. For rank-0 SCFTs, one can fix the overall phase factor ambiguity of $\CH(\vec{z}_\alpha)$ by requiring that
\begin{align}
 \CH(\vec{z}_\alpha ; M=0 , \nu=\pm 1) \in \mathbb{R}_{>0} \textrm{ for all } \vec{z}_\alpha \in \CS_{\rm BE}\;. \label{phase factor of H}
\end{align} 
This is possible since all the  $\CH(\vec{z}_\alpha ; M=0 , \nu=\pm 1)$ have the same phase  for rank-0 SCFTs.

For the $\widetilde{\CT}_{(2,4r)}$ theory, there are $r$ Bethe-vacua, $\{\vec{z}_{\alpha}\}_{\alpha=0}^{r-1}$, and their handle-gluing/fibering operators in the A-twisting limit, $M\rightarrow 0$ and $\nu \rightarrow -1$, are
\begin{align}
\begin{split}
&\CH(\vec{z}_\alpha; M=0, \nu=-1) = (S_{(1,1),(1,2\alpha+1)})^{-2}\;,
\\
&\CF(\vec{z}_\alpha; M=0, \nu=-1)^2 =  e^{2\pi i \delta} \exp \left(4\pi i h_{(s=1,t=2\alpha+1)}\right)\;. \label{evidence 1 for SM(2,4r)-I}
\end{split}
\end{align}
with a $\delta \in \mathbb{Q}$. Here $S$ and $h_{(s,t)}$ is the S-matrix \eqref{S-matrix} and conformal dimensions \eqref{c and h} of $SM(2,4r)$. 

Let $L$ be the supersymmetric Wilson loop of gauge charge $\vec{\mathfrak{q}}  = (\mathfrak{q}_1, \ldots, \mathfrak{q}_r)$. The twisted partition function with insertion of the loop operator can be computed as
\begin{align}
&\mathcal{Z}^L_{\CM_{g, p \in 2\mathbb{Z}}} (M, \nu) = \sum_{\vec{z}_\alpha \in \CS_{\rm BE}} \CH(\vec{z}_\a)^{g-1}\CF (\vec{z}_\a)^p L (\vec{z}_\alpha)\;,
\\
&\textrm{with } L(\vec{z}) := \prod_{a=1}^r (z_a)^{\mathfrak{q}_a}\;.
\end{align}
For the $\widetilde{\CT}_{(2,4r)}$ theory, let $\{L_\alpha\}_{\alpha=0,1,\ldots, r-1 }$ be supersymmetric Wilson loop operators  with following gauge charge $\vec{\mathfrak{q}}_\alpha$:
\begin{align}
\vec{\mathfrak{q}}_{ \alpha} = (0,1,2,\ldots, \alpha-1 , \alpha,\ldots, \alpha )\;. \label{Loops for T(2,4r)}
\end{align}
Then, one can check that
\begin{align}
W_\b (\a) := L_\alpha (\vec{z}_\beta) = \frac{S_{(1,2\a+1), (1, 2\b+1)}}{S_{(1,1),(1,2\b+1)}}\;. \label{evidence 2 for SM(2,4r)-I}
\end{align}
Using the handle-gluing and  fibering operators, the round 3-sphere in \eqref{S^3_b (K,Q)} can be written in the following Bethe-sum \cite{Closset:2017zgf,Gang:2019jut}:
\begin{align}
&|\CZ^{(K,Q)}_{S^3_{b=1}} (\vec{M}, \vec{\nu})| = |\sum_{\vec{z}_\alpha\in \CS_{\rm BE}} \CH(\vec{z}_\alpha;\vec{M}, \vec{\nu})^{-1}\CF (\vec{z}_\alpha;\vec{M},\vec{\nu})|\;, \label{Bethe-sum of ZS3}
\end{align}
only when the $\vec{\nu}$  satisfies following conditions \cite{Gang:2019jut}\footnote{To compute the round 3-sphere partition function, one should use the $(\CH,\CF)$ in the spin-structure choice $\nu_R = 0 (\textrm{mod 1})$, which are generally different from our $(\CH, \CF)$ computed in the $\nu_R = \frac{1}2  (\textrm{mod 1})$. The $(\CH,\CF)$  are independent on the spin-structure choices when the $\vec{\nu}$ satisfies the condition.  }
\begin{align}
K_{aa} + \nu_a \in 2 \mathbb{Z}\label{condition on nu}
\end{align}
For the $\widetilde{\CT}_{(2,4r)}$ theory, the condition is met if
\begin{align}
\nu_1 \in 2\mathbb{Z}+1, \quad \nu_{a>1} \in 2\mathbb{Z}\;.
\end{align}
For general $\vec{\nu}$, the round 3-sphere partition function can be computed using the following relation\footnote{Note that the squashed 3-sphere partition function depends only on the holomorphic combinations $\vec{M}+(i \pi +\frac{\hbar}2) \vec{\nu}$.}
\begin{align}
\CZ^{(K,Q)}_{S^3_{b=1}} (\vec{M}, \vec{\nu}) = |\CZ^{(K,Q)}_{S^3_{b=1}} (\vec{M}+2\pi i (\vec{\nu}-\vec{\nu}_0), \vec{\nu}_0)|\;, \label{Bethe-sum T(2,4r)}
\end{align}
with $\vec{\nu}_0$ is chosen to satisfy the condition in \eqref{condition on nu}.

\paragraph{Half-indices} The half index of the $\widetilde{\CT}_{(2,4r)}$ theory with $\mathbb{B}= (\CD, D_c)$ boundary condition is  \cite{Dimofte:2017tpi}
\begin{align}
\begin{split}
I_{\rm half}(q;\eta, \nu) & = I^{(K,Q)}_{\rm half} (q; \vec{\eta}, \vec{\nu} = (\nu-1)\vec{A})\big{|}_{\eta_a \rightarrow \eta^{A_a}}
\\
&= \sum_{\mathbf{m} \in (\mathbb{Z}_{\geq 0})^r } \frac{q^{\frac{1}2 \mathbf{m}\cdot K \cdot \mathbf{m} } (\eta (-q^{1/2})^{(\nu-1)})^{- \vec{A} \cdot \mathbf{m}}}{(q)_{m_1} \ldots (q)_{m_r}}\;. \label{half-index (2,4r)}
\end{split}
\end{align}
where
\begin{align}
\begin{split}
I^{(K,Q)}_{\rm half}(q;\vec{\eta}, \vec{\nu}) &= \frac{1}{(q)^r_\infty} \sum_{\mathbf{m}\in \mathbb{Z}^r} q^{\frac{1}2 \mathbf{m}\cdot K\cdot \mathbf{m}} \prod_{a=1}^r \left( (\eta_a(-q^{1/2})^{\nu_a})^{m_a} (q^{1-Q_a m_a};q)_\infty \right)
\\
&= \sum_{\mathbf{m} \in (\mathbb{Z}_{\geq 0})^r } \frac{q^{\frac{1}2 \mathbf{m}\cdot K \cdot \mathbf{m} } \prod_{a=1}^r (\eta_a (-q^{1/2})^{\nu_a})^{- m_a}}{(q)_{Q_1 m_1} \ldots (q)_{Q_r m_r}}\;. \label{half-index (K,Q)}
\end{split}
\end{align}
Here $\mathbb{B}=(\CD, D_c)$ is  Dirichlet boundary condition ($\CD$) for vector multiplets and  deformed Dirichlet boundary condition ($D_c$) for  chiral multiplet.  In terms of 2D (0,2) subalgebra, a 3D chiral multiplet is decomposed into a 2d chiral multiplet $\Phi_{2d}$ and a 2d Fermi multiplet. The deformed Dirichlet boundary condition $D_c$(or Dirichlet boundary condition $D$) is imposing $\Phi_{2d}=c$ (or $\Phi_{2d}=0$) with non-zero $c$. The boundary condition breaks all the $U(1)^r$ gauge symmetry while preserving the $U(1)$ R-symmetry. We assign zero R-charge to the chiral fields.

The half-index with the Wilson loop operator $L_\alpha$  in \eqref{Loops for T(2,4r)} becomes \cite{Dimofte:2017tpi}
\begin{align}
I^{L_\alpha}_{\rm half} (q; \eta, \nu) =  \sum_{\mathbf{m} \in (\mathbb{Z}_{\geq 0})^r } \frac{q^{\frac{1}2 \mathbf{m}\cdot K \cdot \mathbf{m} } (\eta (-q^{1/2})^{(\nu-1)})^{- \vec{A} \cdot \mathbf{m}}q^{- \vec{\mathfrak{q}}_\alpha \cdot \mathbf{m}}}{(q)_{m_1} \ldots (q)_{m_r}}
\end{align}
In the A-twisting limit, $\eta \rightarrow 1$ and $\nu \rightarrow -1$, the half-indices reproduce the characters of 2D RCFT $(SM(2,4r))\otimes (\textrm{free Fermion})$
\begin{align}
\begin{split}
&q^{h_\alpha - \frac{c}{24}} I^{L_\alpha}_{\rm half} (q; \eta=1, \nu=-1) =   q^{h_\alpha - \frac{c}{24}}  \sum_{\mathbf{m} \in (\mathbb{Z}_{\geq 0})^r } \frac{q^{\frac{1}2 \mathbf{m}\cdot K \cdot \mathbf{m} +\vec{A}\cdot \mathbf{m}-\vec{\mathfrak{q}}_\alpha \cdot \mathbf{m} } }{(q)_{m_1} \ldots (q)_{m_r}}
\\
&=\left( \chi_{s=1, t= 2\alpha +1} (q)  \textrm{ of $SM(2,4r)$ in \eqref{SM-characters}}\right)  \times  \chi_{F} (q) \;,  \label{evidence 3 for SM(2,4r)-I}
\end{split}
\end{align}
up to an overall factor $q^{h_\alpha - \frac{c}{24}}$ where
\begin{align}
c= 8 - \frac{3}{2r } - 6 r\;, \quad h_\alpha = \frac{\alpha (1 - 2 r + \alpha)}{4 r}\;. \label{c and h of SM(2,4r)*(free F)}
\end{align}
The $c$ is the central charge of the product RCFT, $\left( c \textrm{ in \eqref{c and h} for $SM(2,4r)$} \right)+\frac{1}2$, and $h_\alpha$ is the conformal dimension of $SM(2,4r)$.
It provides a novel fermionic sum expression for the product RCFT characters. 
In the above, $\chi_{F} (q)$ is the character of 2D free Majorana-Weyl fermion theory, whose bulk 3D TQFT is the Ising spin-TQFT $SO(1)_1$, 
\begin{align}
\chi_{F}(q) = q^{- \frac{1}{48}}\mathbb{I}_D (x=-q^{1/2};q) =q^{- \frac{1}{48}} \prod_{n=0}^{\infty} (1+q^{n+1/2}) = q^{- \frac{1}{48}}(-q^{1/2};q)_{\infty}\;. \label{free-fermion character}
\end{align}
We define
\begin{align}
\mathbb{I}_D (x;q) :=(x^{-1}q;q)_{\infty} = \prod_{n=0}^{\infty} (1-q^{n+1}x^{-1})\;, \label{half-index:ID}
\end{align}
which is the half-index for the $\CT_\Delta$ theory  in the Dirichlet boundary ($D$) and with $R(\Phi)=0$. For general R-charge, the half index for $\CT_\Delta$ is $\mathbb{I}_D(x (-q^{1/2})^{R(\Phi)};q)$. 

The equations in \eqref{evidence 1 for SM(2,4r)-I},\eqref{evidence 2 for SM(2,4r)-I},\eqref{evidence 3 for SM(2,4r)-I} and the Figure \ref{fig:F-maximization1} give non-trivial evidences for the proposal  in \eqref{proposal for SM(2,4r)} for the $\widetilde{\CT}_{(2,4r)}$.  They are all compatible with the bulk-boundary dictionaries listed in Table \ref{Table : Dictionaries}.

\subsubsection{UV description II  } \label{sec : UV for SM(2,4r)-II}
The 2nd UV gauge theory description is (for $r=2,4,6,\ldots$)
\begin{align}
\begin{split}
&\CT_{(2,2r+4) }:= \frac{(\CT_\Delta)^{\otimes r}}{[U(1)^{r}_{\mathbf Q}]_{K} }\textrm{ with superpotential } \CW = \CO_{(\mathbf{m}_1, \mathbf{n}_1)}+\ldots +\CO_{(\mathbf{m}_{r-1}, \mathbf{n}_{r-1})}
\\
& K= C(T_r)^{-1}:= \begin{pmatrix}
1&1&1&\cdots&1&1 \\
1&2&2&\cdots&2&2 \\
1&2&3&\cdots&3&3 \\
\vdots&\vdots&\vdots&\ddots&\vdots&\vdots\\
1&2&3&\cdots&r-1& r-1\\
1&2&3&\cdots&r-1&r
\end{pmatrix}, \quad \mathbf{Q} = \mathbb{I}_{r\times r},
\\
&\mathbf{m}_{1} = (1,1,-1,\mathbf{0}_{r-3}),\;\mathbf{m}_{2\leq I\leq r-2}=(\mathbf{0}_{I-2},-1,1,1,-1,\mathbf{0}_{r-I-2}),\;\mathbf{m}_{r-1}=(\mathbf{0}_{r-2},-2,2),
\\
&\mathbf{n}_{1\leq I\leq r-1} = (\mathbf{0}_{r})\;. \label{UV theory for (2,4r)-II}
\end{split}
\end{align}
Here the $C(T_r)$ is the Cartan matrix of the tadpole graph, obtained by folding the Cartan matrix of $A_r$ in half.
The $\CN=2$ gauge theory has a $U(1)$ flavor symmetry, $U(1)_A$, whose charge is given as
\begin{align}
A = \vec{A}\cdot \vec{T}=\sum_{a=1}^{r}\left[ \frac{a+1}{2}\right]T_{a}= 
T_{1}+T_{2}+2T_{3}+2T_{4}+\ldots+\frac{r}{2}T_{r-1}+\frac{r}{2}T_{r}\;.
\label{A for SM(2,2r+4)}
\end{align}
In the above equation square brackets mean integer part. Solving the constraints in \eqref{constraints by superpotential}, the $(\vec{M}, \vec{\nu})$ can be parameterized as
\begin{align}
\vec{M} = \vec{A} M\;, \quad \vec{\nu} = (\nu-1) \vec{A}\;. 
\label{SM(2,2r+4) parametrization}
\end{align}
The bulk supersymmetric partition functions of the gauge theory can be computed as 
\begin{align}
\begin{split}
&\CZ_{S^3_b} (b^2, M , \nu) = \CZ^{(K,Q)}_{S^3_b} (\vec{M} = \vec{A}M, \vec{\nu} = (\nu-1)\vec{A})\;,
\\
& \CI_{\rm sci} (q;\eta, \nu) = \CI^{(K,Q)}_{\rm sci} \left(q;\vec{\eta}, \vec{\nu} = (\nu-1) \vec{A}\right)\bigg{|}_{\eta_a \rightarrow \eta^{A_a}}\;,
\\
& \CZ_{\CM_{g, p\in 2\mathbb{Z}}} (M, \nu) = \CZ^{(K,Q)}_{\CM_{g, p\in 2\mathbb{Z}}} (\vec{M} = \vec{A}M, \vec{\nu} = (\nu-1)\vec{A})\;.
\label{SM(2,2r+4) partition functions}
\end{split}
\end{align}
where the $\CZ_{S^3_b}^{(K,Q)},\CI_{\rm sci}^{(K,Q)}$  and $ \CZ_{\CM_{g, p\in 2\mathbb{Z}}} $ are given in \eqref{S^3_b (K,Q)}, \eqref{SCI (K,Q)} and \eqref{twisted-ptn-(K,Q)} respectively. Superconformal R-charge can be determined by F-maximization, in the same way as \eqref{eq:fmax} and one can confirm that the $\nu=0$ in the parametrization \eqref{SM(2,2r+4) parametrization}  corresponds to the superconformal R-charge. Numerically, one can check that
\begin{align}
\left( \CZ_{S^3_b} (b^2=1, M, \nu) \textrm{ of } \CT_{(2,2r+4)}\right) = \left( \CZ_{S^3_b} (b^2=1, M, \nu) \textrm{ of } \widetilde{T}_{(2,2r+4)} \textrm{ in \eqref{UV theory for (2,4r)-I}}\right)\;.
\end{align}
For the superconformal index we find that, for all $r\in 2\mathbb{Z}_{\geq 1}$
\begin{align}
\begin{split}
&\CI_{\rm sci} (q;\eta, \nu=0) =1-q^{1/2} - \left(1+\eta +\frac{1}\eta\right)q-\left(2+\eta +\frac{1}\eta\right)q^{3/2}+\ldots\;,
\\
&\CI_{\rm sci} (\eta =1, \nu=\pm 1) = 1\;,
\\
& \left( \CI_{\rm sci} (\eta =1, \nu=\pm 1) \textrm{ of $\CT_{(2,2r+4)}$} \right) =  \left( \CI_{\rm sci} (\eta =1, \nu=\pm 1) \textrm{ of $\widetilde{\CT}_{(2,2r+4)}$ in \eqref{UV theory for (2,4r)-I}} \right)\;.
\end{split}
\end{align}
As argued for the $\widetilde{\CT}_{(2,2r+4)}$ case, the superconformal index give non-trivial evidences for the SUSY enhancement, $\CN=2 \rightarrow \CN=5$, in the IR.
 
In the twisted partition computation, there are $(\frac{r}{2}+1)$ Bethe-vacua, $\{\vec{z}_{\alpha}\}_{\alpha=0}^{\frac{r}{2}}$, and their handle-gluing/fibering operators in the A-twisting limit, $\vec{M}\rightarrow \vec{0}$ and $\vec{\nu} \rightarrow -2 \vec{A}$, are
\begin{align}
\begin{split}
&\CH(\vec{z}_\alpha; \vec{M}=0, \vec{\nu}=-2\vec{A}) = (S_{(1,1),(1,2\alpha+1)})^{-2}\;,
\\
& \left(\CF(\vec{z}_\alpha; \vec{M}=0, \vec{\nu}=-2\vec{A}) \right)^2 =  e^{2\pi i \delta} \exp \left(4\pi i h_{(s=1,t=2\alpha+1)}\right)\;, \label{evidence 1 for SM(2,4r)-II}
\end{split}
\end{align}
with a $\delta \in \mathbb{Q}$. Here $S$ and $h_{(s,t)}$ is the S-matrix \eqref{S-matrix} and conformal dimensions \eqref{c and h} of $SM(2,2r+4)$.

The half-index for the $\CT_{(2,2r+4)}$ theory is
\begin{align}
\begin{split}
I_{\rm half}(q;\eta, \nu) & = I^{(K,Q)}_{\rm half} (q; \vec{\eta}, \vec{\nu} = (\nu-1)\vec{A})\big{|}_{\eta_a \rightarrow \eta^{A_a}}
\\
&= \sum_{\mathbf{m} \in (\mathbb{Z}_{\geq 0})^r } \frac{q^{\frac{1}2 \mathbf{m}\cdot K \cdot \mathbf{m} } (\eta (-q^{1/2})^{(\nu-1)})^{- \vec{A} \cdot \mathbf{m}}}{(q)_{\mathbf{m}_1} \ldots (q)_{\mathbf{m}_r}}\;. \label{half-index (2,2r+4)}
\end{split}
\end{align}
Let $\{L_\alpha\}_{\alpha=0,1,\ldots, \frac{r}{2} }$ be supersymmetric Wilson loop operators  with gauge charge $\vec{\mathfrak{q}}_\alpha$:
\begin{align}
\vec{\mathfrak{q}}_{ \alpha} = (1,1,2,2,\ldots,\alpha -1,\alpha -1,\alpha,\alpha,\alpha,\ldots,\alpha)\;.
\end{align}
The half-index with the loop operator $L_\alpha$ becomes
\begin{align}
I^{L_\alpha}_{\rm half} (q; \eta, \nu) =  \sum_{\mathbf{m} \in (\mathbb{Z}_{\geq 0})^r } \frac{q^{\frac{1}2 \mathbf{m}\cdot K \cdot \mathbf{m} } (\eta (-q^{1/2})^{(\nu-1)})^{- \vec{A} \cdot \mathbf{m}}q^{- \vec{\mathfrak{q}}_\alpha \cdot \mathbf{m}}}{(q)_{m_1} \ldots (q)_{m_r}}\;.
\end{align}
In the A-twisting limit, $\eta \rightarrow 1$ and $\nu \rightarrow -1$, the half-indices reproduce the characters of 2D RCFT $SM(2,2r+4)$
\begin{align}
\begin{split}
&q^{h_\alpha - \frac{c}{24}} I^{L_\alpha}_{\rm half} (q; \eta=1, \nu=-1) =   q^{h_\alpha - \frac{c}{24}}  \sum_{\mathbf{m} \in (\mathbb{Z}_{\geq 0})^r } \frac{q^{\frac{1}2 \mathbf{m}\cdot K \cdot \mathbf{m} +\vec{A}\cdot \mathbf{m}-\vec{\mathfrak{q}}_\alpha \cdot \mathbf{m} } }{(q)_{m_1} \ldots (q)_{m_r}}
\\
&= \left( \chi_{(s=1, t= 2\alpha +1)} (q)  \textrm{ of $SM(2,2r+4)$ in \eqref{SM-characters}} \right)\;.  \label{half index for T(2,2r+4)}
\end{split}
\end{align}
Here $K$ and $A$ are given in \eqref{UV theory for (2,4r)-II} and \eqref{A for SM(2,2r+4)}, and the $c$ and $h_{\alpha}$ are the central charge and the conformal dimension of $SM(2,2r+4)$ as given in \eqref{c and h}. The half-indices in \eqref{half index for T(2,2r+4)} reproduce the known fermionic sum expression for the characters of $SM(2,2r+4)$ in \cite{Melzer:1994qp}. 

The nice matches in \eqref{evidence 1 for SM(2,4r)-II} and  \eqref{half index for T(2,2r+4)} support the proposal  in \eqref{proposal for SM(2,4r)} for the $\CT_{(2,4r)}$ case. 

\subsubsection{UV description III  } \label{sec: UV III}
The 3rd UV description is
\begin{align}
\begin{split}
&\widetilde{\CT}'_{(2, 4r)} :=  SU(2)_{k=r}^{\frac{1}2 \oplus \frac{1}2 }
\\
&:=\textrm{ $SU(2)$ gauge theory coupled to a half hypermultiplet and a half twisted}
\\
& \quad \;\;\textrm{-hypermultiplet in the fundamental representations with Chern-Simons level $k=r$.} \label{UV theory for (2,4r)-III}
\end{split}
\end{align}
The theory  is a  rank-0 $\CN=4$ SCFT and actually has $\CN=5$ supersymmetry  \cite{Hosomichi:2008jb,Gang:2021hrd}. 
\begin{table}[ht]
	\begin{center}
		\begin{tabular}{|c|c|c|c|}
			\hline
			Chiral multiplet  &  $SU(2)$ & $R_{\nu=0}$ &  $A$
			\\
			\hline \hline
			$\Phi_1$ & $\mathbf{2}$ & $\frac{1}2$ & $+\frac{1}2$
			\\
			\hline
			$\Phi_2$ & $\mathbf{2}$ & $\frac{1}2$ & $-\frac{1}2$
			\\
			\hline
				\end{tabular}
	\end{center}
	\caption{Matter contents of $SU(2)_{k}^{\frac{1}2\oplus \frac{1}2}$ theory  in terms of $\CN=2$ chiral multiplets. $SU(2)$, $R_{\nu=0}$ and $A$ represent $SU(2)$ gauge symmetry, superconformal R-charge and charge of $U(1)_A$ symmetry respectively.  The theory has a superpotential term $\CW \propto \frac{1}{k} (\epsilon_{ij}\Phi^i_1 \Phi^{j}_2)^2 $ where $i$ and $j=1,2$ are indices for $\mathbf{2}$ of $SU(2)$ gauge group. }
	\label{SU(2)+1/2+1/2}
\end{table}

\paragraph{Squashed 3-sphere partition function} The partition function is given as
\begin{align}
\begin{split}
\CZ_{S^3_b} (M, \nu) &= \int \frac{dZ}{\sqrt{2\pi \hbar}} \CI_\hbar (Z,M, \nu) \textrm{ with}
\\
\CI_\hbar (Z, M, \nu) &=2 \sinh(Z) \sinh (\frac{2\pi i Z}{\hbar}) \exp \left(\frac{(r+1) Z^2}{\hbar}\right)
\\
& \times  \prod_{\epsilon_1, \epsilon_2 \in \{\pm1\}} \psi_\hbar \left( \epsilon_1 Z+\epsilon_2 \frac{M+\nu (i\pi +\hbar/2)}{2} + \frac{(i \pi +\hbar/2)}2\right)\;.
\end{split}
\end{align}
Here $(M,\nu)$ is the (rescaled real mass, R-symmetry mixing parameter) for the $U(1)_A$ symmetry. 
%
%
%
%
%
\paragraph{Superconformal index} The superconformal index is
\begin{align}
\begin{split}
\CI_{\rm sci} (q;\eta, \nu) = &\sum_{m \in \mathbb{Z}_{\geq 0 }} \oint_{|u|=1}\frac{du}{2\pi i u} \Delta (m,u)u^{2(r+1)m}\CI_\Delta \left(m,u (\eta q)^{1/2} \right)  
\\
& \times  \CI_\Delta \left(-m,u^{-1} (\eta q)^{1/2} \right)  \CI_\Delta \left(m,u \eta^{-1/2} \right)   \CI_\Delta \left(-m,u^{-1} \eta^{-1/2} \right)\big{|}_{\eta \rightarrow \eta (-q^{1/2})^{\nu-1}}\;. 
\end{split} \label{SCO T(2,4r)-2}
\end{align}
Here $\Delta (m,u)$ is the contribution from the $SU(2)$ vector multiplet
\begin{align}
\begin{split}
&\Delta (m,u) = \frac{1}{{\rm Sym}(m)} (q^{\frac{m}2}u-q^{-\frac{m}2}u^{-1})(q^{\frac{m}2}u^{-1}-q^{-\frac{m}2}u)\;,
\\
&\textrm{where }\textrm{Sym}(m) := \begin{cases}
2, \quad m=0
\\
1, \quad m>0\;.
\end{cases}
\end{split}
\end{align}
One can check that the index matches with the superconformal index \eqref{SCO T(2,4r)-1} computed using the UV gauge theory  in \eqref{UV theory for (2,4r)-I}. 

\paragraph{Twisted partition functions}  In the asymptotic $\hbar\rightarrow 0$ limit, the integrand of squashed 3-sphere partition function behaves as (in the limit $\nu\rightarrow -1$)
\begin{align}
\begin{split}
&\log \CI_\hbar (Z, M, \nu) \xrightarrow{\quad \hbar \rightarrow 0 \quad } \frac{1}{\hbar}\CW_0 (Z;M,\nu) +\CW_1(Z;M, \nu)\;,
\\
&\CW_0 = \pm 2\pi i Z+(r+1)Z^2+ \sum_{\epsilon_2, \epsilon_2 \in \{\pm 1\}} {\rm Li}_2 (\epsilon_2 e^{-\epsilon_1 Z-\epsilon_2 \frac{M}{2}})\;,
\\
&\CW_1 = - \frac{1}2   \log (1-e^{-Z-\frac{M}2}) -\frac{1}2 \log (1-e^{Z-\frac{M}2})  +\log (\sinh Z)\;.
\end{split}
\end{align}
The corresponding Bethe equation is
\begin{align}
P(z;M, \nu=-1) = \exp (\partial_Z W_0)|_{Z\rightarrow \log z} = \frac{z^{2 r} \left(e^{M/2}+z\right) \left(e^{M/2} z-1\right)}{\left(e^{M/2}-z\right) \left(e^{M/2} z+1\right)}=1\;.
\end{align}
In the A-twisting limit, $(M,\nu)=(0, -1)$, the $P(z) = -z^{2r}$ and there are $r$ Bethe-vacua 
\begin{align}
\begin{split}
&\CS_{\rm BE}(M=0, \nu=-1) = \{z \;:\; P(z, M=0,\nu=-1)=1 \textrm{ and }z^2 \neq 1 \}/\mathbb{Z}_2 
\\
&= \bigg{\{}z_\alpha = \exp \left(\frac{i \pi (2\alpha+1)}{2r}\right):\alpha=0, \ldots, r-1\bigg{\}}\;,
\end{split}
\end{align}
taking into account  the quotient by the Weyl symmetry $\mathbb{Z}_2$, $z \rightarrow 1/z$. 
The corresponding handle-gluing operator $\CH$ is
\begin{align}
\begin{split}
&\CH (z; M=0, \nu=-1)= -\frac{1}4 (\partial_Z \partial_Z \CW_0) e^{-2\CW_1}\bigg{|}_{Z \rightarrow \log z } = \frac{2 r z}{4 (z+1)^2} \;,
\\
&\Rightarrow \CH(z_\alpha) = \frac{r}{2} \cos \left(\frac{(2\a+1)\pi)}{4r}\right)^{-2} = (S_{(1,1),(1,2\a+1)})^{-2}\;. \label{evidence 1 for SM(2,4r)-III}
\end{split}
\end{align}
The factor $1/4$ comes from $1/|W(G)|^2$ \cite{Gang:2021hrd} and the  $(-1)$ is to fix the overall phase factor of $\CH$ according to \eqref{phase factor of H}.
The corresponding fibering-gluing operator $\CF$ is
\begin{align}
\begin{split}
&\CF (z; M, \nu)= \exp \left(-\frac{\CW_0 - Z\partial_Z W_0-M\partial_M W_0}{2\pi i } \right)\bigg{|}_{Z\rightarrow \log z}\;.
\end{split}
\end{align}
Using the expression, one can confirm that
\begin{align}
|\CF (z_\alpha;M=0, \nu=-1)|=1 \textrm{ and }\left( \frac{\CF (z_{\alpha};M=0, \nu=-1)}{\CF (z_{\alpha=0};M=0, \nu=-1)}  \right)^2= \exp (4\pi i h_{(1,2\alpha+1)})\;.  \label{evidence 2 for SM(2,4r)-III}
\end{align}
\paragraph{Half-indices} The half index for the $SU(2)_{k=r}^{\frac{1}2 \oplus \frac{1}2}$ theory is \footnote{Here $\eta$ is the fugacity for the charge $A'=A+J_3$ where the $J_3$ is the Cartan of the $SU(2)$ gauge group and $R_\nu = R_{\nu=0}+\nu A'$. }
\begin{align}
\begin{split}
&I_{\rm half :\mathbb{B}} (q;\eta, \nu) = \textrm{Tr}_{\CH(HS^2:\mathbb{B})} (-1)^{R_\nu} q^{R_\nu+j_3} x^{2J_3 }\eta^{A+J_3}\big{|}_{x\rightarrow 1}
\\
&= \sum_{m\in \mathbb{Z}} \frac{q^{(r-1)m^2}x^{2(r-1)m} \eta^{(r-1)m}}{(q;q)_\infty (q^{1+2m}x^2 \eta ;q)_{\infty} (q^{1-2m}x^{-2} \eta^{-1};q)_{\infty}}
\\
&\times  \mathbb{I}_D \left(-q^{m+1/2}  x\eta  ;q \right) \mathbb{I}_D \left(-q^{-m+1/2} x^{-1} ;q\right) \mathbb{I}_D \left(q^{m} x ;q\right) \mathbb{I}_D \left(q^{-m} (x\eta )^{-1};q\right) \bigg{|}_{\eta  \rightarrow \eta (-q^{1/2})^{\nu-1}, x\rightarrow 1}\;.
\end{split} \label{half-index (2,4r)-2}
\end{align}
The half index agrees with the half index of  the $\widetilde{\CT}_{(2,4r)}$ theory in \eqref{UV theory for (2,4r)-I}:
\begin{align}
\left(I_{\rm half} (q;\eta, \nu) \textrm{ in \eqref{half-index (2,4r)-2}}\right) = \left(I_{\rm half} (q;\eta, \nu) \textrm{ in \eqref{half-index (2,4r)}} \right)\;.
\end{align}
In the A-twisting limit, $\eta\rightarrow 1$ and $\nu\rightarrow-1$, the half-index is  related to the vacuum character of $SM(2,4r)$  as follows
\begin{align}
q^{-\frac{c}{24}}I_{\rm half :\mathbb{B}} (q;\eta=1, \nu=-1) = \left( \chi_{(1,1)}(q)  \textrm{ of $SM(2,4r)$ in \eqref{SM-characters}}\right)\times  \left( \chi_F (q) \textrm{ in \eqref{free-fermion character}}\right)\;, \label{evidence 3 for SM(2,4r)-III}
\end{align}
where $c$ is given in \eqref{c and h of SM(2,4r)*(free F)}.
In the limit, we impose the Dirichlet boundary condition for  $SU(2)$ vector multiplet. We assign R-charge $R=(-1,1,0,2)$ to the 4 ($\mathbf{2}+\mathbf{2}$) chiral multiplets and  impose Dirichlet boundary condition for the 3 chirals with $R\neq 0$ and deformed Dirichlet boundary condition for the chiral with $R=0$. The R-charge assignment and boundary condition break the $SU(2)$ gauge symmetry while preserving the R-symmetry.

The nice matches in \eqref{evidence 1 for SM(2,4r)-III},\eqref{evidence 2 for SM(2,4r)-III} and \eqref{evidence 3 for SM(2,4r)-III} support the proposal  in \eqref{proposal for SM(2,4r)} for the $\widetilde{\CT}'_{(2,4r)}$.
 
\subsection{Bulk dual rank-0 SCFT of $SM(3, 6r-5)$}
 In the following we propose UV gauge theory description of $\mathcal{T}_{(3,6r-5)}$ modulo a decoupled  Ising spin-TQFT, $SO(1)_1$:
 \begin{align}
 &\left( \widetilde{\CT}_{(3,6r-5)} \textrm{ in \eqref{UV theory for (3,6r-5)}} \right) \xrightarrow{\;\textrm{A-twisting/bulk-boundary}\;} SM(3,6r-5) \otimes (\textrm{Free  fermion}). \label{proposal for SM(3,6r-5)}
 \end{align}
\subsubsection{UV description   }
The UV gauge theory description is ($r\geq 2$),
\begin{align}
\begin{split}
&\widetilde{\CT}_{(3,6r-5) }:= \frac{(\CT_\Delta)^{\otimes r}}{[U(1)^{r}_{\mathbf Q}]_{K} }\textrm{ with superpotential } \CW = \CO_{(\mathbf{m}_1, \mathbf{n}_1)}+\ldots +\CO_{(\mathbf{m}_{r-1}, \mathbf{n}_{r-1})}
\\
& K= \begin{pmatrix}
1&-1&-1&\cdots&-1&-1 \\
-1&2&2&\cdots&2&2 \\
-1&2&4&\cdots&4&4 \\
\vdots&\vdots&\vdots&\ddots&\vdots&\vdots\\
-1&2&4&\cdots&2(r-2)&2(r-2)\\
-1&2&4&\cdots&2(r-2)&2(r-1)
\end{pmatrix}, \quad \mathbf{Q} =\textrm{diag} \{ \mathbf{1}_{r-1}, 2\}\;,
\\
&\mathbf{m}_1 = (2,\mathbf{0}_{r-1}), \; \mathbf{m}_2 = (0,2,-1,\mathbf{0}_{r-3}), \; \mathbf{m}_{3 \leq I \leq r-1} = (\mathbf{0}_{I-2},-1,2,-1,\mathbf{0}_{r-I-1}) \; ,
\\
&\mathbf{n}_1 = (0,\mathbf{2}_{r-2},1), \; \mathbf{n}_2 = (1,\mathbf{0}_{r-1}), \; \mathbf{n}_{3 \leq I \leq r-1} = (\mathbf{0}_{r}) \; .
\end{split} \label{UV theory for (3,6r-5)}
\end{align}
 The $\CN=2$ gauge theory has a $U(1)$ flavor symmetry, $U(1)_A$, whose charge is given as
\begin{align}
A = \vec{A}\cdot \vec{T}=  \sum_{a=1}(a-1)T_a  = T_2 + 2T_3+\ldots +(r-1)T_r \;. \label{A for (3,6r-5)}
\end{align}
Solving the constraints in \eqref{constraints by superpotential}, the $(\vec{M}, \vec{\nu})$ can be parameterized as
\begin{align}
\vec{M} = \vec{A} M\;, \quad \vec{\nu} = (\nu-1) \vec{A}\;. 
\end{align}
The bulk supersymmetric partition functions of the gauge theory can be computed as 
\begin{align}
\begin{split}
&\CZ_{S^3_b} (b^2, M , \nu) = \CZ^{(K,Q)}_{S^3_b} (\vec{M} = \vec{A}M, \vec{\nu} = (\nu-1)\vec{A})\;,
\\
& \CI_{\rm sci} (q;\eta, \nu) = \CI^{(K,Q)}_{\rm sci} \left(q;\vec{\eta}, \vec{\nu} = (\nu-1) \vec{A}\right)\bigg{|}_{\eta_a \rightarrow \eta^{A_a}}\;,
\\
& \CZ_{\CM_{g, p\in 2\mathbb{Z}}} (M, \nu) = \CZ^{(K,Q)}_{\CM_{g, p\in 2\mathbb{Z}}} (\vec{M} = \vec{A}M, \vec{\nu} = (\nu-1)\vec{A})\;.
\end{split}
\end{align}
where the $\CZ_{S^3_b}^{(K,Q)},\CI_{\rm sci}^{(K,Q)}$  and $ \CZ_{\CM_{g, p\in 2\mathbb{Z}}} $ are given in \eqref{S^3_b (K,Q)}, \eqref{SCI (K,Q)} and \eqref{twisted-ptn-(K,Q)} respectively. Superconformal R-charge can be determined by F-maximization, in the same way as \eqref{eq:fmax}. 
\begin{figure}[h]
	\centering
	\includegraphics[width=0.495\textwidth]{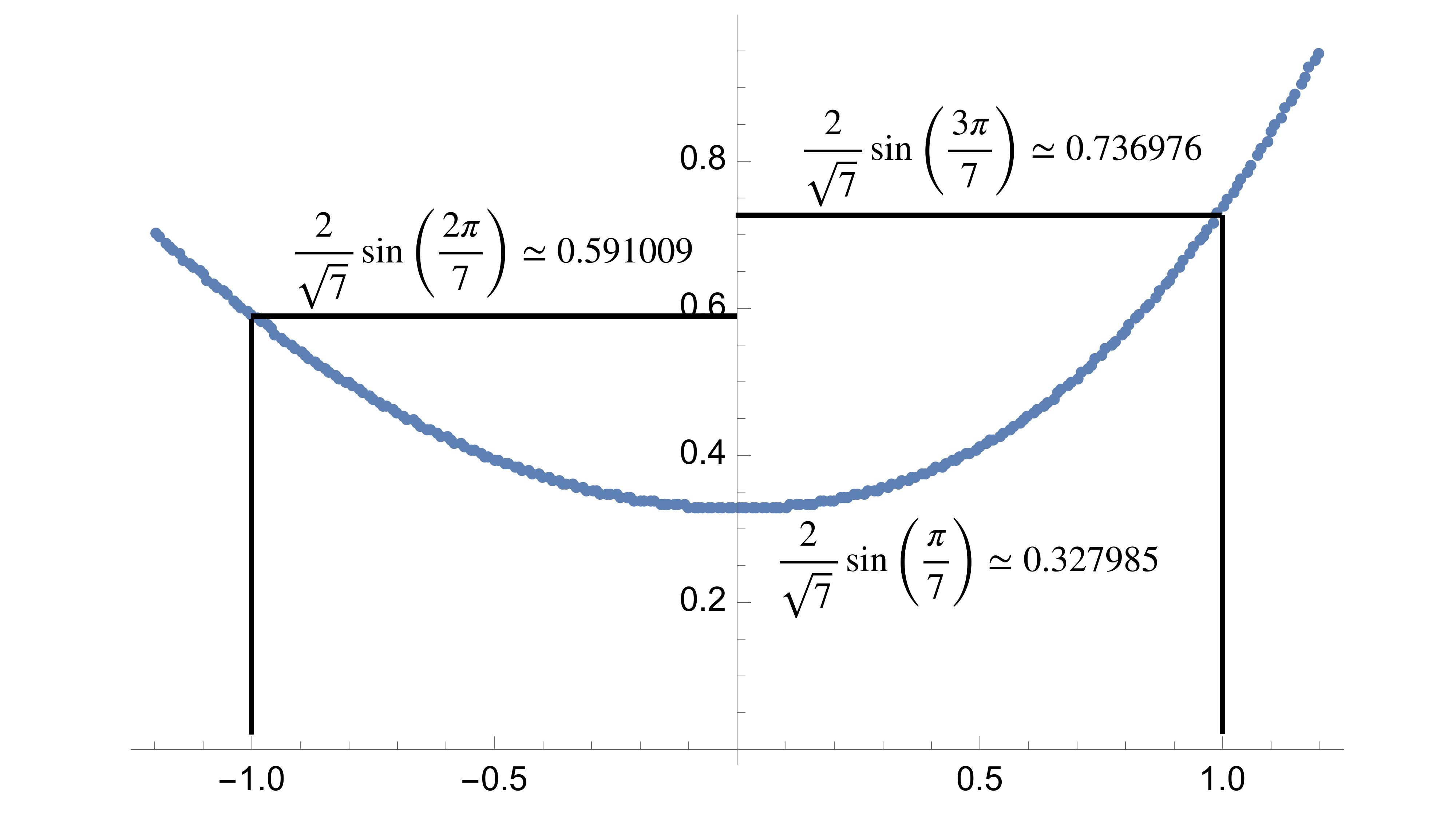}
	\includegraphics[width=0.495\textwidth]{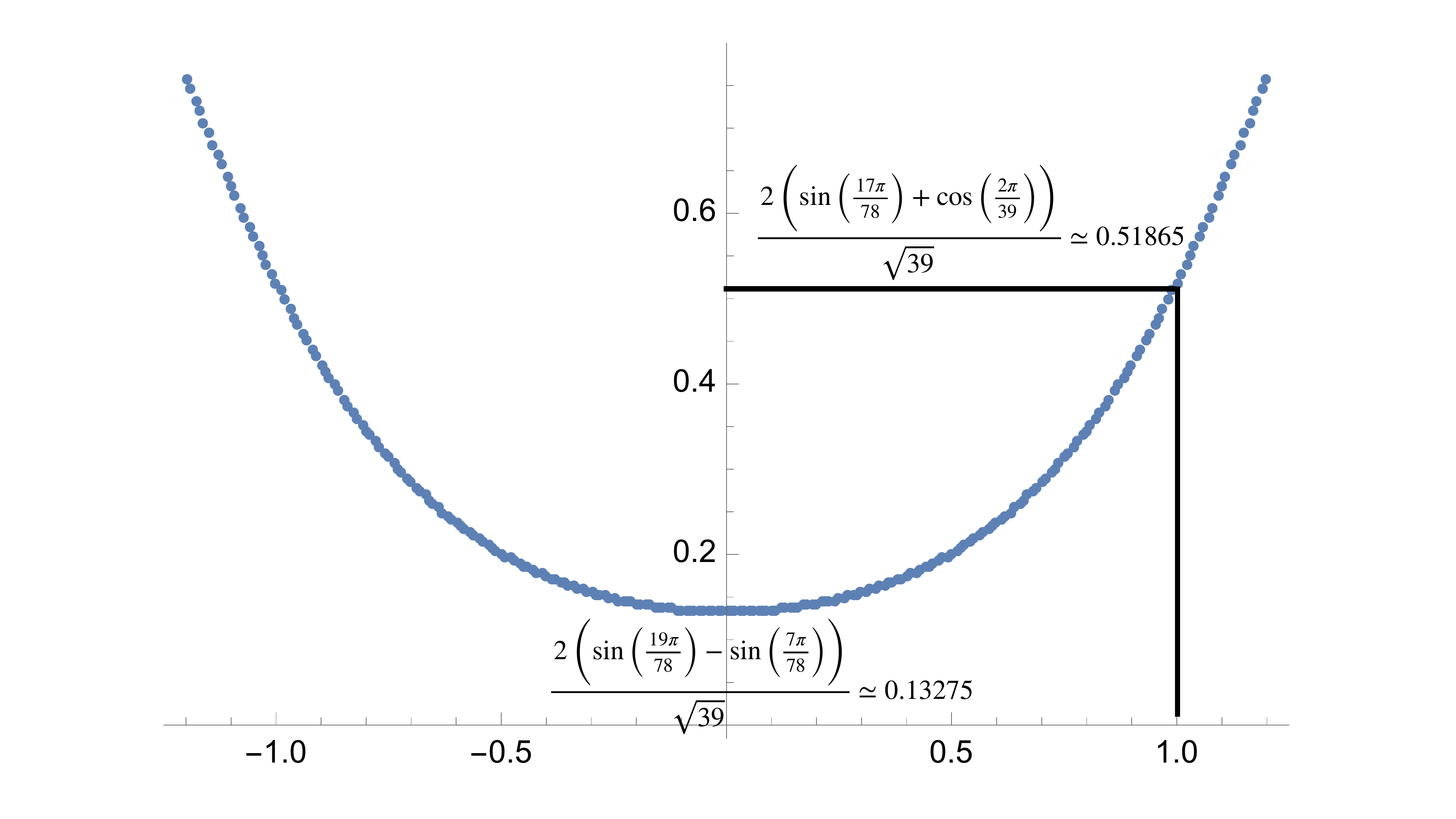}
	\caption{Graph of $|Z_{S^3_b} (b^2=1, M=0, \nu)|$ for $\widetilde{\CT}_{(3,7)}$ theory (Left) and for $\widetilde{\CT}_{(3,13)}$ theory (Right).  For both cases, the round 3-sphere partition functions are minimized at $\nu=0$. The values of  $|Z_{S^3_b} (b^2=1, M=0, \nu=- 1)|$ are identical to the $|S_{\alpha=0, \beta=0} = S_{(1,1),(1,1)}|$ of $SM(3,7)$ and $SM(3,13)$ respectively.   The values at $\nu=0$ and $\nu=1$ of $\widetilde{\CT}_{(3,7)}$ are  $| S_{(1,1),(1,3)}|$ and $| S_{(1,1),(1,5)}|$ of $SM(3,7)$. The values at $\nu=0$  of $\widetilde{\CT}_{(3,13)}$ are  $| S_{(1,1),(1,5)}|$ of $SM(3,13)$.  }
	\label{fig:F-maximization2}
\end{figure}
For the superconformal index we find that
\begin{align}
\begin{split}
&r=2:
\\
& 1-q-\left(\eta +\frac{1}{\eta }\right) q^{3/2}-2 q^2-\eta  q^{5/2}+\left(\frac{1}{\eta ^2}-1\right) q^3+\left(\frac{1}{\eta }-\eta \right) q^{7/2}+\frac{q^4}{\eta ^2}
\\
&+\left(\eta +\frac{3}{\eta }\right) q^{9/2}+\left(\eta ^2+\frac{2}{\eta ^2}+4\right) q^5+\left(3 \eta +\frac{5}{\eta }\right) q^{11/2}+\left(\eta ^2+\frac{1}{\eta ^2}+6\right) q^6+\ldots
\\
&r=3:
\\
&1-q-\left(\eta +\frac{1}{\eta }\right) q^{3/2}-q^2+\left(2 \eta +\frac{2}{\eta }\right) q^{5/2}+\left(2 \eta ^2+\frac{2}{\eta ^2}+6\right) q^3+\left(6 \eta +\frac{6}{\eta }\right) q^{7/2}
\\
&+\left(\eta ^2+\frac{1}{\eta ^2}+8\right) q^4+\left(-\eta ^3-\frac{1}{\eta ^3}+5 \eta +\frac{5}{\eta }\right) q^{9/2}+\left(-\eta ^2-\frac{1}{\eta ^2}+7\right) q^5+\ldots
\\
&\textrm{for all $r\geq 2$:}
\\
&\CI_{\rm sci} (q;\eta, \nu=0) =1-q-\left(\eta+\frac{1}{\eta}\right)q^{\frac{3}{2}} +\ldots\;,
\\
&\CI_{\rm sci} (\eta =1, \nu=\pm 1) = 1\;.
\end{split}
\end{align}
The index shows non-trivial evidences for the SUSY enhancement, $\CN=2 \rightarrow \CN=4$, in the IR.  First, only $q^{\mathbb{Z}/4}$-terms (actually only $q^{\mathbb{Z}/2}$) appears in the index which is compatible with the fact that $\frac{R_{\nu=0}}2 + j_3\in \frac{\mathbb{Z}}4 $ for any $\CN\geq 3 $ theory. This is quite non-trivial fact since the superconformal R-charge $R_{\nu=0}$ is  determined by extremizing the non-trivial function $F(\nu) = \log |\CZ_{S^3_b} (b^2=1, M=0, \nu)|$ as drawn in Figure \ref{fig:F-maximization2}. The index also contains  contributions from two extra-SUSY current multiplets, which are $-(\eta+\frac{1}\eta)q^{1/2}$. The last properties of the index imply that the IR SCFT are of rank-0 if the SUSY enhancement really occurs. 

In the twisted partition computation, there are $(3r-3)$ Bethe-vacua, $\{\vec{z}_{\alpha}\}_{\alpha=0}^{3r-4}$, and their handle-gluing/fibering operators in the A-twisting limit, $\vec{M}\rightarrow \vec{0}$ and $\vec{\nu} \rightarrow -2 \vec{A}$, are
\begin{align}
\begin{split}
&\CH(\vec{z}_\alpha; \vec{M}=0, \vec{\nu}=-2\vec{A}) = (S_{(1,1),(1,2\alpha+1)})^{-2}\;,
\\
& \left(\CF(\vec{z}_\alpha; \vec{M}=0, \vec{\nu}=-2\vec{A}) \right)^2 =  e^{2\pi i \delta} \exp \left(4\pi i h_{(s=1,t=2\alpha+1)}\right)\;. \label{evidence 1 for SM(3,6r-5)}
\end{split}
\end{align}
with a $\delta \in \mathbb{Q}$. Here $S$ and $h_{(s,t)}$ are the S-matrix \eqref{S-matrix} and conformal dimensions \eqref{c and h} of $SM(3,6r-5)$ respectively.

The half-index for the $\widetilde{\CT}_{(3,6r-5)}$ theory is
\begin{align}
\begin{split}
I_{\rm half}(q;\eta, \nu) & = I^{(K,Q)}_{\rm half} (q; \vec{\eta}, \vec{\nu} = (\nu-1)\vec{A})\big{|}_{\eta_a \rightarrow \eta^{A_a}}
\\
&= \sum_{\mathbf{m} \in (\mathbb{Z}_{\geq 0})^r } \frac{q^{\frac{1}2 \mathbf{m}\cdot K \cdot \mathbf{m} } (\eta (-q^{1/2})^{(\nu-1)})^{- \vec{A} \cdot \mathbf{m}}}{(q)_{\mathbf{m}_1} \ldots (q)_{\mathbf{m}_{r-1}} (q)_{2\mathbf{m}_r}}\;. \label{half-index (3,6r-5)}
\end{split}
\end{align}
The half-index in the A-twisting limit, $\eta\rightarrow 1$ and $\nu \rightarrow -1$, reproduces the vacuum character the supersymmetric minimal model $SM(3,6r-5)$ up to a decoupled free-fermion character
\begin{align}
\begin{split}
& I_{\rm half} (q;\eta=1, \nu=-1) = \sum_{\mathbf{m} \in (\mathbb{Z}_{\geq 0})^r } \frac{q^{\frac{1}2 \mathbf{m}\cdot K \cdot \mathbf{m} }q^{\vec{A} \cdot \mathbf{m}}}{(q)_{\mathbf{m}_1} \ldots (q)_{\mathbf{m}_{r-1}} (q)_{2\mathbf{m}_r}}\;
\\
&=q^{\frac{c}{24}+ \frac{1}{48}}  \times \left(  \chi_{(1,1)} (q) \textrm{ of }SM(3,5r-7) \textrm{ in \eqref{SM-characters}}\right) \times \left( \chi_F (q) \textrm{ in \eqref{free-fermion character}}\right)\;. \label{evidence 2 for SM(3,6r-5)}
\end{split}
\end{align}
Here $K$ and $A$ are given in \eqref{UV theory for (3,6r-5)} and \eqref{A for (3,6r-5)}  and the $c$ is the central charge of $SM(2,6r-5)$ as given in \eqref{c and h}.

The non-trivial confirmations in \eqref{evidence 1 for SM(3,6r-5)} and \eqref{evidence 2 for SM(3,6r-5)} of the bulk-boundary dictionaries  listed in the Table \ref{Table : Dictionaries} strongly support the proposal in \eqref{proposal for SM(3,6r-5)}.

\subsection{Bulk dual rank-0 SCFT  of $SM(3, 6r-7)$}
Here we propose UV gauge theory description of $\mathcal{T}_{(3,6r-7)}$ modulo a decoupled  spin Ising TQFT, $SO(1)_1$:
\begin{align}
&\left(\widetilde{ \CT}_{(3,6r-7)} \textrm{ in \eqref{UV theory for (3,6r-7)}} \right) \xrightarrow{\;\textrm{A-twisting/bulk-boundary}\;} SM(3,6r-7) \otimes (\textrm{Free  fermion}). \label{proposal for SM(3,6r-7)}
\end{align}
\subsubsection{UV description  }
The UV gauge theory description is ($r\geq 3$),\footnote{For $r=2$ with $K=\begin{pmatrix} 1 & -1 \\ -1 & 4 \end{pmatrix}$, the theory $ \frac{(\CT_\Delta)^{\otimes r}}{[U(1)^{r}_{\mathbf{Q}}]_{K} }$  have two independent $1/2$ BPS monopole operators, $\CO_1 = \Phi_2 V_{(1,0)}$ and $\CO_2 =\Phi_1 V_{(0,1)} $.  After the superpotential deformation with $\CW =(\CO_1)^2 + \CO_2$, the theory has a mass gap and flows to a unitary TQFT which is a bulk dual of  the  $SM(3,5)\otimes (\textrm{free Majornara fermion})$. }
\begin{align}
\begin{split}
&\widetilde{\CT}_{(3,6r-7) }:= \frac{(\CT_\Delta)^{\otimes r}}{[U(1)^{r}_{\mathbb{Q}}]_{K} }\textrm{ with superpotential } \CW = \CO_{(\mathbf{m}_1, \mathbf{n}_1)}+\ldots +\CO_{(\mathbf{m}_{r-1}, \mathbf{n}_{r-1})}
\\
& K= \begin{pmatrix}
1&-1&-1&\cdots&-1&-1 \\
-1&2&2&\cdots&2&2 \\
-1&2&\ldots&\cdots&\ldots &\ldots \\
\vdots&\vdots&\vdots&\ddots&\vdots&\vdots\\
-1&2&\ldots &\cdots&2(r-2)&2(r-2)\\
-1&2 &\ldots&\cdots&2(r-2)&2r
\end{pmatrix}, \quad \mathbf{Q} =\textrm{diag} \{ \mathbf{1}_{r-1}, 2\}\;,
\\
&\mathbf{m}_1 = (2,\mathbf{0}_{r-1}), \; \mathbf{m}_2 = (0,2,-1,\mathbf{0}_{r-3}), \; \mathbf{m}_{3 \leq I \leq r-2} = (\mathbf{0}_{I-2},-1,2,-1,\mathbf{0}_{r-I-1}) \; ,
\\
& \mathbf{m}_{r-1} = (\mathbf{0}_{r-2},-1,1) \;,
\\
&\mathbf{n}_1 = (0,\mathbf{2}_{r-2},1), \; \mathbf{n}_2 = (1,\mathbf{0}_{r-1}), \; \mathbf{n}_{3 \leq I \leq r-1} = (\mathbf{0}_{r}) \; .
\end{split} \label{UV theory for (3,6r-7)}
\end{align}
Notice that the $K$ matrix is identical to the $K$ in \eqref{UV theory for (3,6r-5)} expcept the last $(r,r)$-component. 
The $\mathcal{N}=2$ gauge theory has a $U(1)_A$ flavor symmetry with charge 
\begin{align}
A = \vec{A}\cdot \vec{T}=  \sum_{a=1}^{r-1}(a-1)T_a +(r-2)T_{r} \;. \label{A for (3,6r-7)}
\end{align}
Solving the constraints in \eqref{constraints by superpotential}, the $(\vec{M}, \vec{\nu})$ can be parameterized as
\begin{align}
\vec{M} = \vec{A} M\;, \quad \vec{\nu} = (\nu-1) \vec{A}\;. 
\end{align}
The bulk supersymmetric partition functions of the gauge theory can be computed as 
\begin{align}
\begin{split}
&\CZ_{S^3_b} (b^2, M , \nu) = \CZ^{(K,Q)}_{S^3_b} (\vec{M} = \vec{A}M, \vec{\nu} = (\nu-1)\vec{A})\;,
\\
& \CI_{\rm sci} (q;\eta, \nu) = \CI^{(K,Q)}_{\rm sci} \left(q;\vec{\eta}, \vec{\nu} = (\nu-1) \vec{A}\right)\bigg{|}_{\eta_a \rightarrow \eta^{A_a}}\;,
\\
& \CZ_{\CM_{g, p\in 2\mathbb{Z}}} (M, \nu) = \CZ^{(K,Q)}_{\CM_{g, p\in 2\mathbb{Z}}} (\vec{M} = \vec{A}M, \vec{\nu} = (\nu-1)\vec{A})\;.
\end{split}
\end{align}
where the $\CZ_{S^3_b}^{(K,Q)},\CI_{\rm sci}^{(K,Q)}$  and $ \CZ_{\CM_{g, p\in 2\mathbb{Z}}} $ are given in \eqref{S^3_b (K,Q)}, \eqref{SCI (K,Q)} and \eqref{twisted-ptn-(K,Q)} respectively. Superconformal R-charge can be determined by F-maximization, in the same way as \eqref{eq:fmax}. 
\begin{figure}[h]
	\centering
	\includegraphics[width=0.6\textwidth]{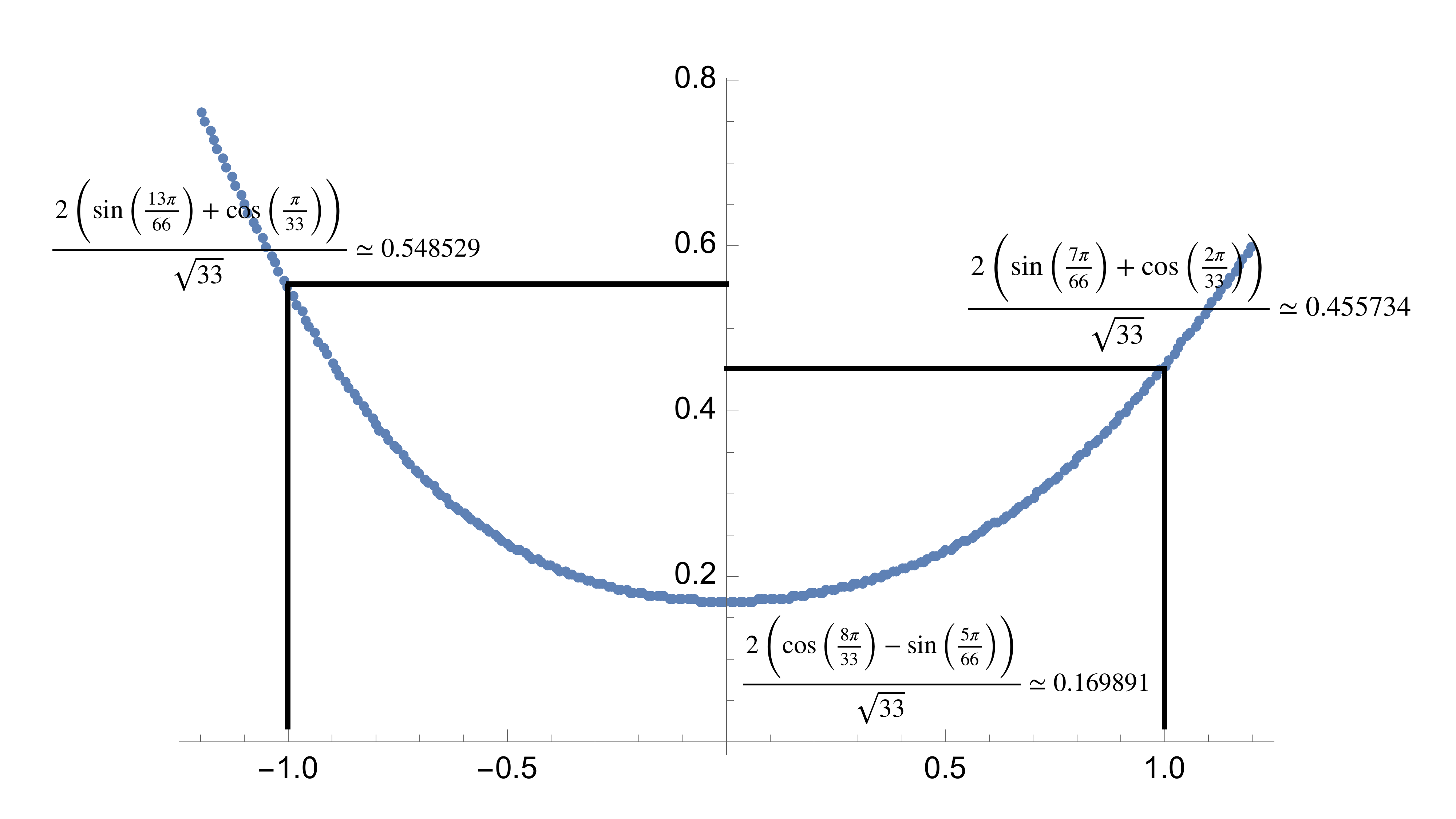}
	\caption{Graph of $|Z_{S^3_b} (b^2=1, M=0, \nu)|$ for $\widetilde{\CT}_{(3,11)}$ theory.  The round 3-sphere partition function is minimized at $\nu=0$. The values at $\nu=-1$, $\nu=0$ and $\nu=1$ are identical to the $|S_{(1,1),(1,1)}|$,  $|S_{(1,1),(1,3)}|$ and  $|S_{(1,1),(1,9)}|$ of  $SM(3,11)$ respectively.   }
	\label{fig:F-maximization3}
\end{figure}
 For the superconformal index we find that
\begin{align}
\begin{split}
&r=3:
\\
& 1-q+\left(-\eta -\frac{1}{\eta }\right) q^{3/2}-2 q^2+\left(\eta ^2+\frac{1}{\eta ^2}+1\right) q^3+\left(2 \eta +\frac{2}{\eta }\right) q^{7/2}+\left(\frac{1}{\eta ^2}+3\right) q^4
\\
&+\left(-\eta ^3+3 \eta +\frac{5}{\eta }\right) q^{9/2}+\left(\frac{3}{\eta ^2}+8\right) q^5+\left(-\eta ^3+3 \eta +\frac{7}{\eta }\right) q^{11/2}+\ldots
\\
&\textrm{For all $r\geq 3$}:
\\
&\CI_{\rm sci} (q;\eta, \nu=0) =1-q-\left(\eta+\frac{1}{\eta}\right)q^{\frac{3}{2}} +\ldots\;,
\\
&\CI_{\rm sci} (\eta =1, \nu=\pm 1) = 1\;.
\end{split}
\end{align}
The index also shows  evidences for the SUSY enhancement, $\CN=2 \rightarrow \CN=4$,  and being of rank 0 SCFT in the IR. 

In the twisted partition computation, there are $(3r-4)$ Bethe-vacua, $\{\vec{z}_{\alpha}\}_{\alpha=0}^{3r-5}$, and their handle-gluing/fibering operators in the A-twisting limit, $\vec{M}\rightarrow \vec{0}$ and $\vec{\nu} \rightarrow -2 \vec{A}$, are
\begin{align}
\begin{split}
&\CH(\vec{z}_\alpha; \vec{M}=0, \vec{\nu}=\vec{0}) = (S_{(1,1),(1,2\alpha+1)})^{-2}\;,
\\
& \left(\CF(\vec{z}_\alpha; \vec{M}=0, \vec{\nu}=\vec{0}) \right)^2 =  e^{2\pi i \delta} \exp \left(4\pi i h_{(s=1,t=2\alpha+1)}\right)\;. \label{evidence 1 for SM(3,6r-7)}
\end{split}
\end{align}
with a $\delta \in \mathbb{Q}$. Here $S$ and $h_{(s,t)}$ is the S-matrix \eqref{S-matrix} and conformal dimensions \eqref{c and h} of $SM(3,6r-7)$.

The half-index for the $\widetilde{\CT}_{(3,6r-7)}$ theory is
\begin{align}
\begin{split}
I_{\rm half}(q;\eta, \nu) & = I^{(K,Q)}_{\rm half} (q; \vec{\eta}, \vec{\nu} = (\nu-1)\vec{A})\big{|}_{\eta_a \rightarrow \eta^{A_a}}
\\
&= \sum_{\mathbf{m} \in (\mathbb{Z}_{\geq 0})^r } \frac{q^{\frac{1}2 \mathbf{m}\cdot K \cdot \mathbf{m} } (\eta (-q^{1/2})^{(\nu-1)})^{- \vec{A} \cdot \mathbf{m}}}{(q)_{\mathbf{m}_1} \ldots (q)_{\mathbf{m}_{r-1}} (q)_{2\mathbf{m}_r}}\;. \label{half-index (3,6r-7)}
\end{split}
\end{align}
The half-index in the A-twisting limit, $\eta\rightarrow 1$ and $\nu \rightarrow -1$, reproduces the vacuum character the supersymmetric minimal model $SM(3,6r-7)$ up to an decoupled free-fermion character
\begin{align}
\begin{split}
& I_{\rm half} (q;\eta=1, \nu=-1) = \sum_{\mathbf{m} \in (\mathbb{Z}_{\geq 0})^r } \frac{q^{\frac{1}2 \mathbf{m}\cdot K \cdot \mathbf{m}+ \vec{A}\cdot \mathbf{m} }}{(q)_{m_1} \ldots (q)_{m_{r-1}} (q)_{2m_r}}\;
\\
&=q^{\frac{c}{24}+ \frac{1}{48}}  \times \left(  \chi_{(1,1)} (q) \textrm{ of }SM(3,6r-7) \textrm{ in \eqref{SM-characters}}\right) \times \left( \chi_F (q) \textrm{ in \eqref{free-fermion character}}\right)\;. \label{evidence 2 for SM(3,6r-7)}
\end{split}
\end{align}
Here the matrix $K$ and the vector $\vec{A}$ is given in \eqref{UV theory for (3,6r-7)} and \eqref{A for (3,6r-7)} respectively.

The non-trivial identities in \eqref{evidence 1 for SM(3,6r-7)} and \eqref{evidence 2 for SM(3,6r-7)} strongly support the proposal in \eqref{proposal for SM(3,6r-7)}.
\section{Summary and Future directions}
In this paper, we propose bulk 3D $\CN=4$ rank-0 theories which are related to the $\CN=1$ supersymmetric minimal models $SM(2,\cdot)$ and $SM(3,\cdot)$ via the bulk-boundary correspondence.  Like most rank-0 SCFTs, the SCFTs are realized as IR fixed points of UV gauge theories with less supersymmetry ($\CN=2$) except for the examples in section \ref{sec: UV III}  which have manifest $\CN=5$ supersymmetry. We support the proposal by checking the dictionaries in the Table \ref{Table : Dictionaries} with explicit computations of supersymmetric partition functions.

It would be interesting to generalize our work to other examples of RCFTs and see if there are some non-unitary RCFTs which can not be realized from 3D rank-0 SCFTs. The strongest form of the non-unitary bulk-boundary correspondence can be stated as that every non-unitary chiral RCFTs can be realized as boundary algebras of 3D rank-0 SCFTs. If true, it would be a crucial advantage of 3D non-unitary bulk-boundary correspondence over (4D SCFTs)/(2D VOAs) correspondence in which only a few classes of non-unitary rational algebra can be realized \cite{Song:2017oew,Beem:2017ooy,Arakawa:2017fdq,Xie:2019vzr}.  So far, we could  find bulk dual rank-0 SCFTs only for some examples of  minimal models, $M(p,q)$ with $p=2$, and supersymmetric minimal models, $SM(p,p')$ with $p=2$ and $p=3$. What about other cases? In the upcoming papers \cite{Gang:2024tlp,BGK}, bulk field theories for  general minimal models and supersymmetric  minimal models are proposed using the 3D-3D correspondence \cite{Terashima:2011qi,Dimofte:2011ju,Gang:2018wek,Cho:2020ljj,Choi:2022dju,Bonetti:2024cvq}.

In this paper, the supersymmetric minimal models arise from topological A-twisting of the bulk rank-0 SCFTs.  It would be interesting to study the boundary chiral algebras of the topologically B-twisted theories and see how the two chiral algebras from A- and B- twistings are related to each other. For the $M(2,q)$ case, the relation was recently studied in \cite{Ferrari:2023fez}. 

\acknowledgments{We would like to thank  Heeyeon Kim, Heesu Kang, Byoungyoon Park, Huijoon Sohn, Spencer Stubbs, Arash Arabi Ardehali and Mykola Dedushenko  for the useful discussion and the collaborations on related topics.
The work of DG and SB is supported in part by the National Research Foundation of Korea grant  NRF-2022R1C1C1011979. DG also acknowledges support by Creative-Pioneering Researchers Program through Seoul National University. }

\appendix
\section{Quantum dilogarithm and tetrahedron index} \label{appendix}
The quantum dilogarithm function $\psi_\hbar (Z)$ ($\hbar :=2\pi i b^2$) is defined by \cite{Faddeev:1993rs} 
\begin{align}
\psi_\hbar (Z):= \begin{cases} 
\prod_{r=1}^{\infty} \frac{1-q^r e^{-Z}}{1-\widetilde{q}^{-r+1} e^{-\widetilde{Z}}}\;, \quad \textrm{if $|q|<1$}
\\
\prod_{r=1}^{\infty} \frac{1-\widetilde{q}^r e^{-\widetilde{Z}}}{1-q^{-r+1} e^{-Z}}\;, \quad \textrm{if $|q|>1$}
\end{cases}
\end{align}
with
\begin{align} 
q=e^{2\pi i b^2}, \quad \widetilde{q}:=e^{2\pi i b^{-2}}, \quad \widetilde{Z}= Z/b^2\;.
\end{align}
The $\psi_\hbar (Z)$ computes the squashed 3-sphere partition function of the $\CT_\Delta$  theory in \eqref{T-Delta} with $Z = M+(i \pi+\frac{\hbar}2 R(\Phi))$, where the $M$ is the rescaled real mass for the $U(1)$ flavor symmetry. 
In the limit $\hbar\rightarrow 0$, the Q.D.L behaves as follows
\begin{align}
\begin{split}
\log \psi_\hbar (Z) \xrightarrow{\;\; \hbar\rightarrow 0 \;\; }& \sum_{n=0}^\infty \frac{B_n \hbar^{n-1}}{n!} {\rm Li}_{2-n} (e^{-Z}) = \frac{1}\hbar {\rm Li}_2 (e^{-Z}) - \frac{1}2 \log (1-e^{-Z})+\ldots. \label{asymptotic of QDL}
\end{split}
\end{align}
Here $B_n$ is the $n$-th Bernoulli number with $B_1=1$. When $b=1$, on the other hand, the function  becomes
\begin{align}
\psi_{\hbar =2\pi i } (Z) = \exp\left( \frac{-(2 \pi +i Z)\log (1-e^{-Z})+i {\rm Li}_2 (e^{-Z})}{2\pi }\right)\;.
\end{align}
The tetrahedron index $\CI_\Delta (m,u)$ is defined as \cite{Dimofte:2011py}
\begin{align}
\begin{split}
&\CI_\Delta (m,u) := \prod_{r=0}^\infty \frac{1-q^{r-\frac{1}2 m+1}u^{-1}}{1-q^{r- \frac{1}2 m} u} =\sum_{e \in \mathbb{Z}}\CI_\Delta^{c} (m, e) u^e,  \;\; 
\\
&\textrm{where  } \CI_\Delta^c (m,e) = \sum_{n=\lfloor e \rfloor }^\infty \frac{(-1)^n q^{\frac{1}2 n (n+1)-(n+\frac{1}2 e)m}}{(q)_n (q)_{n+e}}\;.
\end{split}
\end{align}
It computes the generalized superconformal index \cite{Kapustin:2011jm} of the $\CT_\Delta$ theory with the R-charge choice $R(\Phi)=0$ where $(m,u)$ are (background monopole flux, fugacity) for the $U(1)$ flavor symmetry. At  general $R$-charge choice, the index becomes $\CI_\Delta(m, u(-q^{1/2})^{R(\Phi)})$. 

\section{Bulk  field theories for $M(2,r+2)$ using $K=C(T_r)^{-1}$ with odd $r$}
In  section \ref{sec : UV for SM(2,4r)-II}, we propose a UV gauge theory description of $\CT_{(2,2r+4)}$ using mixed CS level $K=C(T_r)^{-1}$ with even $r$. Here we propose bulk field theories, say $\widetilde{\CT}_{M(2,r+2)}$, for the Virasoro minimal model $M(2,r+2)$ using  $K=C(T_r)^{-1}$ with odd $r$. More precisely, we propose that\footnote{In \cite{Gang:2023rei}, they propose bulk field theory $\CT_{M(2,r+2)}$ for minimal model $M(2,r+2)$ with $r\in 2\mathbb{Z}_{\geq 1}+1$ using $K=2 C(T_{(r-1)/2})^{-1}$. We expect an IR duality between $\left(\CT_{M(2,r+2)} \textrm{ in  \cite{Gang:2023rei}} \right)\otimes SO(1)_1$ and $\left( \widetilde{\CT}_{M(2,r+2)}  \textrm{ in \eqref{UV theory for (2,4r)-II-2}}\right)$. See also \cite{Comi:2023lfm,Gang:2024tlp} for another  dual theories  with manifest $\CN=3$ supersymmetry. }
\begin{align}
\begin{split}
&\textrm{For $r\in 2\mathbb{Z}_{\geq 1} +1$,}
\\
&\widetilde{\CT}_{M(2,r+2)} \xrightarrow{\;\textrm{A-twisting/bulk-boundary}\;} M(2,r+2) \otimes (\textrm{Free  fermion}),
\end{split}
\end{align}
where
\begin{align}
\begin{split}
&\widetilde{\CT}_{M(2,r+2) }:= \frac{(\CT_\Delta)^{\otimes r}}{[U(1)^{r}_{\mathbf Q}]_{K} }\textrm{ with superpotential } \CW = \CO_{(\mathbf{m}_1, \mathbf{n}_1)}+\ldots +\CO_{(\mathbf{m}_{r-1}, \mathbf{n}_{r-1})}\;,
\\
& K= C(T_r)^{-1}:= \begin{pmatrix}
1&1&1&\cdots&1&1 \\
1&2&2&\cdots&2&2 \\
1&2&3&\cdots&3&3 \\
\vdots&\vdots&\vdots&\ddots&\vdots&\vdots\\
1&2&3&\cdots&r-1& r-1\\
1&2&3&\cdots&r-1&r
\end{pmatrix}, \quad \mathbf{Q} = \mathbb{I}_{r\times r},
\\
\begin{split}
&\mathbf{m}_{1} = (1,1,-1,\mathbf{0}_{r-3}),\;\mathbf{m}_{2\leq I\leq r-2}=(\mathbf{0}_{I-2},-1,1,1,-1,\mathbf{0}_{r-I-2}),\;\mathbf{m}_{r-1}=(\mathbf{0}_{r-2},-2,2),
\\
&\mathbf{n}_{1\leq I\leq r-1} = (\mathbf{0}_{r})\;. 
\end{split}\label{UV theory for (2,4r)-II-2}
\end{split}
\end{align}
%
%
%
The $\CN=2$ gauge theory has a $U(1)$ flavor symmetry, $U(1)_A$, whose charge is given as
\begin{align}
A = \vec{A}\cdot \vec{T}=\sum_{a=1}^{r}\left[ \frac{a}{2}\right]T_{a}= 
T_{2}+T_{3}+2T_{4}+2T_{5}+\ldots+\frac{r-1}{2}T_{r-1}+\frac{r-1}{2}T_{r}\;.
\end{align}
The computation of various BPS partition functions can be done  as in the main text.  For the superconformal index we find that
\begin{align}
\begin{split}
&\CI_{\rm sci} (q;\eta, \nu=0) =1-q-\left(\eta+\frac{1}{\eta}\right)q^{\frac{3}{2}}+\ldots\;,
\\
&\CI_{\rm sci} (\eta =1, \nu=\pm 1) = 1\;.
\end{split}
\end{align}
The index shows the evidence for the SUSY enhancement, $\CN=2 \rightarrow \CN=4$, in the IR. In the twisted partition computation, there are $\frac{r+1}{2}$ Bethe-vacua, $\{\vec{z}_{\alpha}\}_{\alpha=0}^{\frac{r-1}{2}}$. Their handle-gluing/fibering operators in the A-twisting limit, $\vec{M}\rightarrow \vec{0}$ and $\vec{\nu} \rightarrow -2 \vec{A}$, are
\begin{align}
\begin{split}
&\CH(\vec{z}_\alpha; \vec{M}=0, \vec{\nu}=-2\vec{A}) = (S_{0, \a})^{-2}\;,
\\
& \left(\CF(\vec{z}_\alpha; \vec{M}=0, \vec{\nu}=-2\vec{A}) \right)^2 =  e^{2\pi i \delta} \exp \left(4\pi i h_{\a}\right)\;.
\end{split}
\end{align}
with a $\delta \in \mathbb{Q}$. Here $S$ and $h_{\a}$ are the S-matrix and conformal dimensions of $M(2,r+2)$:
\begin{align}
\begin{split}
&S_{\a,\b} = \frac{2 (-1)^{\frac{r-1}2+\a+\b}}{\sqrt{r+2}} \sin \left(\frac{2\pi (\a+1)(\b+1)}{r+2} \right)\;,
\\
&h_\alpha = \frac{\alpha (\alpha-r)}{2r+4}\;. \label{S and h of M(2,r+2)}
\end{split}
\end{align}
Let $\{L_\alpha\}_{\alpha=0,1,\ldots, \frac{r-1}{2} }$ be supersymmetric Wilson loop operators  with gauge charge $\vec{\mathfrak{q}}_\alpha$:
\begin{align}
\vec{\mathfrak{q}}_{ \alpha} = (0,1,1,2,2,\ldots,\alpha -1,\alpha -1,\alpha,\alpha,\alpha,\ldots,\alpha)\;.
\end{align}
In the A-twisting limit, $\eta \rightarrow 1$ and $\nu \rightarrow -1$, the half-indices reproduce the characters of 2D RCFT $(M(2,r+2))\otimes (\textrm{free Majonara})$
\begin{align}
\begin{split}
&q^{h_\alpha - \frac{c}{24}} I^{L_\alpha}_{\rm half} (q; \eta=1, \nu=-1) =   q^{h_\alpha - \frac{c}{24}}  \sum_{\mathbf{m} \in (\mathbb{Z}_{\geq 0})^r } \frac{q^{\frac{1}2 \mathbf{m}\cdot K \cdot \mathbf{m} +\vec{A}\cdot \mathbf{m}-\vec{\mathfrak{q}}_\alpha \cdot \mathbf{m} } }{(q)_{m_1} \ldots (q)_{m_r}}
\\
&=\left( \chi_{\a} (q)  \textrm{ of $M(2,r+2)$}\right)  \times\left( \chi_F (q) \textrm{ in \eqref{free-fermion character}}\right)\;,  \label{evidence for M(2,r+2)}
\end{split}
\end{align}
up to an overall factor $q^{h_\alpha - \frac{c}{24}}$ where conformal dimension $h_\alpha$ is given in \eqref{S and h of M(2,r+2)} and the central charge $c$ of the product RCFT is 
\begin{align}
c= \frac{3}{2}-\frac{3r^2}{r+2}\;, \quad h_\alpha = \frac{\alpha (\alpha-r)}{2r+4}\;.
\end{align}
The character $\chi_\alpha (q)$ of $M(2,r+2)$ is  \cite{feigin1983verma,felder1989brst}
\begin{align}
\begin{split}
&\chi_{\alpha}(q) =\frac{q^{-\Delta^{(2,r+2)}_{1,\alpha+1} } }{(q)_\infty} \sum_{k\in \mathbb{Z}} \left( q^{\Delta^{(2,r+2)}_{1+4k,\alpha+1}}-q^{\Delta^{(2,r+2)}_{1+4k,-\alpha-1}}  \right)\;,
\\
&\Delta^{(p,q)}_{r,s} := \frac{(r q-sp)^2-(q-p)^2}{4 pq}\;.
\end{split}
\end{align}
\newpage
\bibliographystyle{ytphys}
\bibliography{ref}

\end{document}